\documentclass[a4paper,fleqn,usenatbib]{mnras}

\usepackage[T1]{fontenc}
\usepackage{ae,aecompl}

\usepackage{graphicx}	
\usepackage{amsmath}	
\usepackage{amssymb}	

\title{Accurate Pre-Eruption and Post-Eruption Orbital Periods for the Dwarf/Classical Nova V1017 Sgr}

\author[I.V. Salazar et al.]{
Irene V. Salazar$^{1}$\thanks{E-mail: ivarga1@lsu.edu}
, Amy LeBleu $^{1}$
, Bradley E. Schaefer $^{1}$
, Arlo U. Landolt $^{1}$
\newauthor and Shawn Dvorak $^{2}$
\\
$^{1}$Department of Physics \& Astronomy, Louisiana State University, Baton Rouge, LA 70803, USA \\
$^{2}$American Association of Variable Star Observers, 49 Bay State Road, Cambridge, MA 02138, USA
}

\date{Accepted XXX. Received YYY; in original form ZZZ}

\pubyear{2016}

\begin{document}
\label{firstpage}
\pagerange{\pageref{firstpage}--\pageref{lastpage}}
\maketitle

\begin{abstract}
V1017 Sgr is a classical nova (in 1919) that displayed an earlier dwarf nova eruption (in 1901), and two more dwarf nova events (in 1973 and 1991).  Previous work on this bright system in quiescence (V=13.5) has only been a few isolated magnitudes, a few spectra, and an ambiguous claim for an orbital period of 5.714 days as based on nine radial velocities.  To test this period, we have collected 2896 magnitudes (plus 53 in the literature) in the UBVRIJHKL bands from 1897 to 2016, making an essentially complete photometric history of this unique cataclysmic variable.  We find that the light curve in all bands is dominated by the ellipsoidal modulations of a G giant companion star, with a post-eruption (after the 1919 nova event) orbital period of 5.786290$\pm$0.000032 days.  This is the longest period for any classical nova, the accretion must be powered by the nuclear evolution of the companion star, and the dwarf nova events occur only because the outer parts of the large disk are cool enough to be unstable.  Furthermore, we measure the pre-eruption orbital period (from 1907 to 1916), and there is a small steady period change in quiescence.  The orbital period has {\it decreased} by 273$\pm$61 parts-per-million across the 1919 eruption, with the significance of the period change being at the 5.7-sigma confidence level.  This is startling and mystifying for nova-theory, because the three known period change effects cannot account for a period decrease in V1017 Sgr, much less one of such a large size.
\end{abstract}




\section{Introduction}

Cataclysmic variables (CVs) are binary systems where an ordinary companion star spills matter via Roche lobe overflow through an accretion disk onto a white dwarf.  CVs display three types of eruptive events.  Classical novae (CNe) are explosive events on white dwarfs in CVs when the accreted mass accumulates to the point where runaway thermonuclear reactions eject mass in an outgoing shell.  Nova eruptions typically last for a year or more and have amplitudes of $>$7 mags.  Dwarf Novae (DNe) events are temporary brightenings of the accretion disk when an instability in the accretion flow causes a sudden surge of mass flowing through the disk.  Dwarf novae events are distinguished by durations of usually days-to-weeks and amplitudes of 1-4 mags.  Symbiotic novae are poorly understood events caused by thermonuclear burning on the white dwarfs in systems that are symbiotic stars (where something like a red giant is orbiting a white dwarf with both hot and cool components in the spectrum).  Symbiotic novae have durations of years-to-decades and amplitudes of 1-6 mags.

V1017 Sgr is an unique CV because it shows two types of eruptive phenomena.  In 1919, it had a nova eruption, with an amplitude of $>$7.2 mags and a total duration of nearly a year.  This event had much too large an amplitude to be a DN event, its total duration is greatly too long to be a DN event, and the light curve shape (with a fading by 3 mag in 125 days) is that of a classical nova.  Dwarf nova events were recorded in 1901, 1973, and 1991, with amplitudes of 3 mags and total durations of a few months each.  These eruptions were spectroscopically confirmed to be dwarf nova events (Vidal \& Rodgers 1974). V1017 Sgr is apparently a symbiotic star, having a reported orbital period of 5.714 days (Sekiguchi 1992) and spectrum of a G-type giant or sub-giant (Kraft 1964; Harrison et al. 1993).

Sekiguchi (1992) claimed that V1017 Sgr has a period of 5.714 days, a period that is greatly larger than almost all CVs. This implies a very large accretion disk where the Roche lobe overflow is driven by the expansion of the companion star due to its nuclear evolution.  The nuclear evolution drives a secular expansion of the companion star that results in a relatively high accretion rate that can support CN events with a reasonable frequency.  The very large accretion disk allows for its outer regions to be sufficiently cool so as to support the dwarf nova instabilities.  Thus, Sekiguchi's long orbital period explains why this one CV system has both CN and DN events.  Only one other system, GK Per, is known to have both CN and DN events, and this also has the long orbital period of 2.0 days.

However, Sekiguchi reported the 5.714 day orbital period with just a set of 9 radial velocity measurements over a 37 day interval of observation.  Any such data set is not robust due to the possibilities of various errors in even one measure.  (A similar case has been the early radial velocity curve for the recurrent nova T CrB, with errors in one or two measures warping the conclusions and driving theorists to claims from which it has taken our field several decades to recover.)  Any such limited data set is also subject to many aliases, and Sekiguchi points to possible periods of 0.851 and 1.212 days.  For the period, he notes that confirmation is needed.  Surprisingly, no follow-up radial velocities have been reported.  Furthermore, since CVs almost always show photometric modulations on the orbital period, Sekiguchi's claimed period could be confirmed with a good light curve showing a photometric modulation on the orbital period.  Yet, no light curves of any type have been published for V1017 Sgr, despite V1017 Sgr being one of the brightest nova in quiescence, at V=13.5 or so.

The orbital period of V1017 Sgr is critical for all physical models, yet Sekiguchi's period is problematic.  In this paper, we report a large collection of magnitudes from 1897 to 2016, and we test to see whether Sekiguchi's period appears as a photometric modulation. 

\section{Spectral Energy Distribution}

A spectral energy distribution (SED) was constructed using data from five sources, see Table 1 and Figure 1. Column one is the band on which the data were taken, column two is the observed flux, $f_{\nu}$, in milliJanskys, column three is the flux adjusted for extinction, $f_{\nu,0}$, in milliJanskys, and column four is the source of the data. Fluxes were corrected for interstellar dust extinction according to the equations of Cardelli, Clayton, and Mathis (1989).  We adopted a value for E(B-V) of 0.39 mag (Webbink et al. 1987).

            The SED is plotted in Figure 1. The variations of flux are consistent with the quiescent levels of the star fluctuating by as much as half a mag.  This is not the classic profile of a CV, which is much brighter in the blue than the red, which decreases according to the power law with $f_{\nu}\propto \lambda^{-1/3}$. Instead, it closely resembles a blackbody with a peak $f_{\nu}$ near 1.0 $\mu$m.  This peak corresponds to a surface temperature near 5200 K, implying a G-type star.  This surface temperature is consistent with the spectral absorption lines for which Kraft (1964) gives a spectral class of G5 IIIp, as confirmed by Sekiguchi (1992).
            
            For the companion star to dominate the SED to this extent, the companion must be significantly large. A large companion forces the period to also be large, so this eliminates the possibility of a period of shorter than a day. Narrowing down the possible period length greatly increases efficiency in running Fourier Transforms with our very long time intervals, as these shorter periods do not have to be investigated. It also eliminated some of Sekiguchi's possible aliases.  

\begin{table}
        \centering
        \caption{Spectral Energy Diagram.}
        \label{tab:table1}
        \begin{tabular}{llll} 
                               \hline
                                Band & $f_{\nu}$ (mJy) & $f_{\nu,0}$ (mJy) & Source\\
                                \hline
B	&	5.7	&	24.6	&	SMARTS	\\
V	&	13.3	&	40.6	&	SMARTS	\\
R	&	21.3	&	49.1	&	SMARTS	\\
I	&	32.1	&	54.7	&	SMARTS	\\
U	&	2.0	&	11.2	&	Landolt (2016)	\\
B	&	5.8	&	25.0	&	Landolt (2016)	\\
V	&	13.7	&	41.6	&	Landolt (2016)	\\
R	&	21.4	&	49.4	&	Landolt (2016)	\\
I	&	32.6	&	55.6	&	Landolt (2016)	\\
J	&	50.6	&	69.3	&	Harrison et al. (1993)	\\
H	&	58.3	&	70.9	&	Harrison et al. (1993)	\\
K	&	42.3	&	47.7	&	Harrison et al. (1993)	\\
J	&	42.2	&	57.8	&	2MASS	\\
H	&	49.4	&	60.1	&	2MASS	\\
K	&	36.0	&	40.6	&	2MASS	\\
WISE1	&	19.5	&	21.3	&	WISE	\\
        \hline
	\end{tabular}
	
\end{table}

\begin{figure}
	\includegraphics[width=1.1\columnwidth]{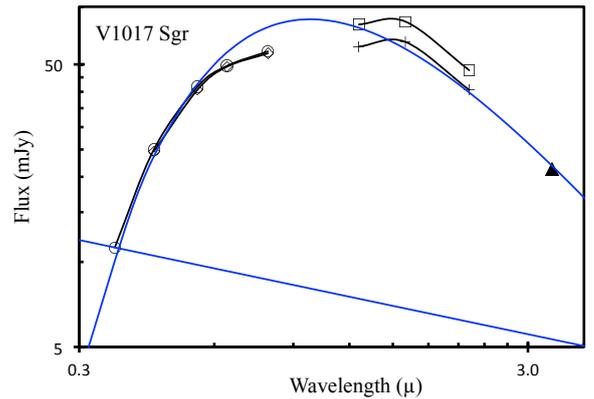}
    \caption{V1017 Sgr spectral energy distribution.  As constructed from Table 1, we see a peak in $f_{\nu,0}$ around 1.0 $\mu$m, which implies that the companion star has a surface temperature of around 5200 K (a G type star).  To guide the eye, the blue curve is a 5200 K blackbody, while the blue diagonal line across the bottom is for the brightest allowable classical accretion disk.  We see that the entire SED for V1017 Sgr is consistent with the 5200 K blackbody.  The companion star dominates over the accretion disk, so the star's surface area must be large, and it must be a giant or a sub-giant.  With this, the orbital period can only be longer than roughly one day so that the companion star can fit into the Roche lobe geometry.}
    \label{fig:Fig. 1}
\end{figure}

\section{Observations}

We have a number of magnitudes from different sources from which we will test V1017 Sgr's period.  We think that these data sets constitute essentially all of the extent photometry of V1017 Sgr.   Details of the observations are specified in Table 2.  An itemized listing of the observations is presented in Table 3.  For the CCD observations and photoelectric photometry measures, the measurement uncertainties were $\sim$0.01 mag, with this being smaller than the usual flickering and variability on V1017 Sgr.

V1017 Sgr has a nearby star that is much brighter; CD -29\degr 15053.  This star is 15.2 arc-seconds almost directly to the east.  The proximity of this star is no problem for CCD photometry for systems with arc-second pixel sizes.  But some systems (the Harvard patrol plates, APASS) have troubles with crowding.  Landolt (2016) reports V=10.183 and B=11.817.  Landolt (2016) also reports 29 UBVRI measures on 23 nights spread over 21 years and finds that the star is constant.

\begin{table*}
	\begin{minipage}{90mm}
	\centering
	\caption{Data Sources.}
	\label{tab:table2}
	\begin{tabular}{lllll} 
		\hline
		Sources & Years & \# & Band & Observer\\
		\hline
HRPO 	&	2015	&	202	&	V	&	Salazar, LeBleu	\\
SMARTS	&	2005 - 2006	&	40	&	BVRI	&	Schaefer	\\
Cerro Tololo 	&	1975 - 2001	&	215	&	UBVRI	&	Landolt	\\
AAVSO	&	1999 - 2015	&	147	&	V	&	Dvorak	\\
AAVSO	&	1955 - 2016	&	2212	&	Visual	&	many	\\
HCO Plates	&	1897 - 1950	&	80	&	B	&	Schaefer	\\
Literature	&	1970 - 1991	&	53	&	UBVRIJHKL	&	$^a$	\\
		\hline
	\end{tabular}
	$^a$ Mumford (1971); Vidal \&  Rodgers (1974); Landolt (1975); Walker (1977); Harrison et al. (1993)
	\end{minipage}
\end{table*}

The first set of data (as listed in Table 2) was taken with the 0.4-m and 0.5-m telescopes at the Highland Road Park Observatory (HRPO) in Baton Rouge, Louisiana. Our primary comparison star is the nearby bright star CD   -29\degr 15053, for which the AAVSO gives a magnitude of V=10.173. This companion star was never saturated in our images, and its light never spilled into the photometry aperture for V1017 Sgr. We obtained 202 good images in the V-band over a span of 10 nights from 23 August to 15 October 2015.  These measures were all obtained within a half-hour interval on each night (although on the second night the light curve covered a 1.7 hour interval), with this being much smaller than any orbital period, so we are only presenting the averaged points for each night.

Our second data set comes from the SMARTS 1.3-m telescope on Cerro Tololo. This data set includes magnitudes in BVRI. The BVRI magnitudes from our three comparison stars were calibrated by A. Landolt. 

The third data set is a long series made on the Cerro Tololo 1.5-m, 1.0-m, 0.9-m, 0.6-m, and 0.4-m telescopes. These observations were all made with photoelectric photometers.  The measurements are essentially simultaneous measures in the UBVRI magnitude system.  The photometry aperture had a diameter of 14 - 17 arc-seconds, and care was taken to slightly offset the center so as to exclude light from CD -29\degr 15053.  A full description of the observations is given in Landolt (2016).  This is the best data set for mapping out the UBVRI colors and for looking at the long term variations in quiescence.

The fourth data set is recorded in the American Association of Variable Star Observers (AAVSO) International Database.  In particular, here, we are only considering the CCD magnitudes recorded with a V-band filter.  S. Dvorak (DKS) was the observer for 139 of these V band magnitudes, while 1 of them was by A. Liu (LAQ) and 7 of them by S. O'Connor (OCN).  The typical 1-sigma photometric uncertainty ranges from 0.02-0.06 mag.  The DKS observations have a large number of nights over a small number of years, and so these data are the best for pulling out any long photometric period.

The fifth data set is the visual observations collected in the AAVSO International Database.  This includes all the measures reported during the DN eruptions as copied in the IAU Circulars.  For the visual observations, the one-sigma uncertainty is roughly 0.2 mag.  We have rejected the magnitude upper-limits as not being useful.  We have also rejected the outliers from several observers, mainly where they were apparently reporting a magnitude for CD -29\degr 15053.  For the magnitudes in quiescence, we see trends with changes on the time scale of five years or so.  These can be characterized as a brightening by $\sim$0.5 mag in the middle 1970s and a fading by $\sim$0.5 mag in the mid-2000s.  Based on our wide experience, we suspect that these are associated with changing comparison star sequences, with the change put in soon after the 1973 DN event and then with the many changes in comparison sequences around 2005 (the `Henden bumps' ).  Further, these changes are not seen in any other data set.  So we conclude that the AAVSO magnitudes in quiescence cannot be used to look for long term trends in the quiescent brightness level of this old nova.  We will use these magnitudes in quiescence for our period searches, but will conclude that these magnitudes have too large a photometric uncertainty, so that the Fourier Transform is very noisy, and thus that the AAVSO quiescent visual magnitudes cannot be used for this purpose either.  The AAVSO visual magnitudes are still very useful, because they are the best measures of the DN events in 1973 and 1991.

The sixth data set includes observations from the archival plates at the Harvard College Observatory (HCO). The HCO plates record sky photographs from the 1890s to 1989 with a wide variety of depths and plate scales. The native system of these plates is essentially the Johnson B system. The comparison stars are nearby stars with their B magnitudes taken from AAVSO.  With this, the resultant derived magnitudes are in the Johnson B system.  The one-sigma uncertainty in the magnitudes is usually from 0.10 to 0.15 magnitudes.  Unfortunately, V1017 Sgr is essentially unresolved from CD -29\degr 15053 on almost all plates.  This means that when the star is in quiescence, the magnitude of the combined star image is not significantly different from the magnitude of CD -29\degr 15053 alone, so there can be no useful information on the brightness of the old nova.  Only for three sets of conditions can we pull out anything useful, but these turn out to be critical for several issues.  (1) Only the A plates have V1017 Sgr well-resolved from CD -29\degr 15053, because they have a very large plate scale.  So only with the A plates can we get B magnitudes in quiescence.  With this, we have 43 magnitudes, 16 from 1897-1916 and 27 from 1923-1950.  These data are critical for evaluating the long term variations in quiescence, including any decline after the eruption as for any sort of hibernation, as well as for looking for changes in the brightness across eruption.  (2) Nova Sagittarii 1919 number 5 was discovered with the Harvard plates (Woods 1919).  All plates that show the nova during its 1919 CN eruption have the nova unresolved from CD -29\degr 15053.  We have 24 magnitude measures, all from 1919, showing the combined light to be significantly brighter than the light from CD -29\degr 15053 alone.  In all cases, we have numerically corrected the combined brightness back to the brightness of V1017 Sgr alone by subtracting the flux of the nearby star.  These data are critical for proving that the 1919 event was a CN eruption, and to establish its basic light curve.  (3) The 1901 DN event is recorded because the combined light gets substantially brighter than the light of CD -29\degr 15053 alone, so the brightness of the nova alone can be pulled out.  We have 13 magnitude measures for the two stars together, all from 1901.  All plates from 1900 and 1902 show the combined light as equaling that of CD -29\degr 15053 alone, so the nova is faint at its quiescent level.  The combined light has been corrected by the subtraction of the constant B=11.817 for CD -29\degr 15053 so as to get the B magnitude of V1017 Sgr alone.  These data are critical for showing the existence of the 1901 event, and for showing that its light curve is the same as the later DN events.  McLaughlin (1946) reported magnitudes from the same set of Harvard plates, but his comparison stars were far from the Johnson B system, his magnitude for CD -29\degr 150503 corrections was off, and he missed all the A plates.  As such, the McLaughlin measures from the Harvard plates are entirely superseded by those in this paper.

The seventh set of data comes from literature (Mumford 1971; Vidal \& Rodgers 1974; Landolt 1975; and Walker 1977; Harrison et al. 1993), with all these being based on photoelectric photometry. These are 53 sparse data points for V1017 Sgr in the bands UBVRIJHKL.  These data are included for completeness, and because they help fill in the long-term time history.

All of the magnitudes from our data sets are listed in Table 3; but with two exceptions.  The first exception to this is the visual magnitudes from the AAVSO.  (These are permanently available freely at the AAVSO web site.)  The second exception is the Landolt magnitudes in the U, R, and I filters, with these being tabulated in Landolt (2016).  In Table 3, magnitudes are listed in the order of the filters (U, B, V, R, I, J, H, K, and L), and then by time within each filter.

\begin{table}
	\centering
	\caption{Magnitudes.}
	\label{tab:table3}
	\begin{tabular}{llllll} 
		\hline
		JD & Year & B. & Mag & Source & Phase\\
		\hline
2440408.6477	&	1969.512	&	U	&	14.99	&	Mumford	&	0.362	\\
2441831.063	&	1973.406	&	U	&	13.2	&	Vidal et al.	&	0.189	\\
2442904.5	&	1976.345	&	U	&	14.86	&	Walker	&	0.703	\\
2414189.5666	&	1897.726	&	B	&	15.0	&	HCO(A2752)	&	0.337	\\
2415526.8690	&	1901.387	&	B	&	13.40	&	HCO(AM798)	&	0.403	\\
2415527.8200	&	1901.390	&	B	&	13.57	&	HCO(AM802)	&	0.567	\\
2415532.7060	&	1901.403	&	B	&	13.45	&	HCO(AM808)	&	0.411	\\
2415533.6970	&	1901.406	&	B	&	13.30	&	HCO(AM817)	&	0.583	\\
2415575.6791	&	1901.520	&	B	&	12.13	&	HCO(AM907)	&	0.836	\\
2415591.5604	&	1901.563	&	B	&	11.60	&	HCO(B27996)	&	0.580	\\
2415604.6146	&	1901.599	&	B	&	11.26	&	HCO(B28160)	&	0.836	\\
2415611.6075	&	1901.618	&	B	&	11.51	&	HCO(B28215)	&	0.044	\\
2415632.5744	&	1901.676	&	B	&	11.04	&	HCO(AM1028)	&	0.667	\\
2415634.5740	&	1901.681	&	B	&	10.56	&	HCO(AM1043)	&	0.013	\\
2415641.5272	&	1901.700	&	B	&	10.61	&	HCO(B28531)	&	0.214	\\
2415661.5540	&	1901.756	&	B	&	10.23	&	HCO(B28621)	&	0.674	\\
2415668.5714	&	1901.774	&	B	&	11.34	&	HCO(B28735)	&	0.887	\\
2417760.8748	&	1907.504	&	B	&	15.5	&	HCO(A8370)	&	0.404	\\
2417805.6689	&	1907.627	&	B	&	15.2	&	HCO(A8446)	&	0.144	\\
2418171.6658	&	1908.629	&	B	&	15.2	&	HCO(A9015)	&	0.382	\\
2418176.6534	&	1908.642	&	B	&	15.1	&	HCO(A9032)	&	0.244	\\
2418197.5656	&	1908.700	&	B	&	15.1	&	HCO(A9066)	&	0.857	\\
2418865.7307	&	1910.529	&	B	&	14.9	&	HCO(A10189)	&	0.306	\\
2418916.5386	&	1910.668	&	B	&	15.3	&	HCO(A10251)	&	0.084	\\
2419677.5367	&	1912.752	&	B	&	15.0	&	HCO(A10615)	&	0.572	\\
2419949.5738	&	1913.496	&	B	&	15.1	&	HCO(A10729)	&	0.576	\\
2419959.8607	&	1913.525	&	B	&	15.0	&	HCO(A10731)	&	0.353	\\
2420002.5895	&	1913.642	&	B	&	14.7	&	HCO(A10772)	&	0.736	\\
2420008.5930	&	1913.658	&	B	&	14.7	&	HCO(A10775)	&	0.773	\\
2420693.7066	&	1915.534	&	B	&	14.9	&	HCO(A11297)	&	0.150	\\
2420724.6561	&	1915.618	&	B	&	15.0	&	HCO(A11393)	&	0.497	\\
2421069.6672	&	1916.563	&	B	&	14.8	&	HCO(A11506)	&	0.109	\\
2422029.9059	&	1919.193	&	B	&	6.43	&	HCO(AC21269)	&	0.023	\\
2422073.8332	&	1919.313	&	B	&	7.62	&	HCO(AC21382)	&	0.615	\\
2422082.8530	&	1919.338	&	B	&	8.03	&	HCO(MF3522)	&	0.173	\\
2422083.8581	&	1919.340	&	B	&	8.47	&	HCO(AM14682)	&	0.347	\\
2422114.8339	&	1919.425	&	B	&	9.15	&	HCO(AM14759)	&	0.701	\\
2422122.7393	&	1919.447	&	B	&	8.91	&	HCO(AC21523)	&	0.067	\\
2422133.6473	&	1919.477	&	B	&	9.07	&	HCO(AM14791)	&	0.952	\\
2422134.6502	&	1919.480	&	B	&	8.68	&	HCO(AM14797)	&	0.126	\\
2422140.7047	&	1919.496	&	B	&	8.80	&	HCO(AC21574)	&	0.172	\\
2422148.6600	&	1919.518	&	B	&	9.15	&	HCO(AC21590)	&	0.547	\\
2422158.6824	&	1919.545	&	B	&	10.06	&	HCO(MF4151)	&	0.279	\\
2422159.6560	&	1919.548	&	B	&	10.06	&	HCO(MF4174)	&	0.447	\\
2422165.6164	&	1919.564	&	B	&	10.61	&	HCO(AC21626)	&	0.477	\\
2422166.5181	&	1919.567	&	B	&	9.72	&	HCO(AM14869)	&	0.633	\\
2422172.6258	&	1919.584	&	B	&	10.16	&	HCO(AM14881)	&	0.689	\\
2422191.5962	&	1919.635	&	B	&	10.52	&	HCO(AM14934)	&	0.968	\\
2422192.5820	&	1919.638	&	B	&	10.58	&	HCO(AM14939)	&	0.138	\\
2422193.5780	&	1919.641	&	B	&	10.94	&	HCO(MF4542)	&	0.310	\\
2422196.5620	&	1919.649	&	B	&	11.15	&	HCO(MF4582)	&	0.826	\\
2422197.5390	&	1919.652	&	B	&	10.94	&	HCO(MF4611)	&	0.995	\\
2422200.5470	&	1919.660	&	B	&	11.15	&	HCO(MF4701)	&	0.514	\\
2422214.5000	&	1919.698	&	B	&	11.78	&	HCO(MC16245)	&	0.926	\\
2422216.5151	&	1919.704	&	B	&	10.94	&	HCO(MF4796)	&	0.274	\\
2422225.5207	&	1919.728	&	B	&	11.21	&	HCO(AM15021)	&	0.831	\\
2423560.6927	&	1923.383	&	B	&	14.5	&	HCO(A12379)	&	0.588	\\
2423575.6624	&	1923.424	&	B	&	14.3	&	HCO(A12414)	&	0.175	\\
2423579.6295	&	1923.435	&	B	&	14.3	&	HCO(A12423)	&	0.861	\\
2423587.6197	&	1923.457	&	B	&	14.7	&	HCO(A12444)	&	0.242	\\
2423603.5608	&	1923.501	&	B	&	14.6	&	HCO(A12483)	&	0.997	\\
2423641.6374	&	1923.605	&	B	&	14.7	&	HCO(A12578)	&	0.577	\\
2423644.6632	&	1923.613	&	B	&	14.3	&	HCO(A12581)	&	0.100	\\
2423906.7383	&	1924.331	&	B	&	14.4	&	HCO(A12917)	&	0.395	\\
		\hline
	\end{tabular}
\end{table}

\begin{table}
	\centering
	\contcaption{Magnitudes}
	\label{tab:continued}
	\begin{tabular}{llllll} 
		\hline
		JD & Year & B. & Mag & Source & Phase\\
		\hline

2425878.2646	&	1929.728	&	B	&	14.7	&	HCO(A14238)	&	0.131	\\
2425881.2833	&	1929.737	&	B	&	14.6	&	HCO(A14240)	&	0.653	\\
2426809.6216	&	1932.278	&	B	&	14.7	&	HCO(A15988)	&	0.096	\\
2429165.2336	&	1938.728	&	B	&	14.5	&	HCO(A20490)	&	0.212	\\
2429165.2446	&	1938.728	&	B	&	14.5	&	HCO(A20491)	&	0.214	\\
2429190.2472	&	1938.796	&	B	&	14.9	&	HCO(A20524)	&	0.535	\\
2429190.2592	&	1938.796	&	B	&	14.8	&	HCO(A20525)	&	0.537	\\
2429191.2421	&	1938.799	&	B	&	14.3	&	HCO(A20533)	&	0.707	\\
2429375.4764	&	1939.303	&	B	&	15.0	&	HCO(A20793)	&	0.547	\\
2429396.4611	&	1939.361	&	B	&	14.7	&	HCO(A20887)	&	0.174	\\
2429396.4791	&	1939.361	&	B	&	14.6	&	HCO(A20888)	&	0.177	\\
2429458.2477	&	1939.530	&	B	&	15.0	&	HCO(A21158)	&	0.853	\\
2429785.3584	&	1940.425	&	B	&	14.9	&	HCO(A21875)	&	0.386	\\
2432441.2277	&	1947.697	&	B	&	14.7	&	HCO(A26000)	&	0.392	\\
2432441.2387	&	1947.697	&	B	&	14.6	&	HCO(A26001)	&	0.394	\\
2433390.6012	&	1950.296	&	B	&	14.6	&	HCO(A27053)	&	0.468	\\
2433455.5168	&	1950.474	&	B	&	14.2	&	HCO(A27246)	&	0.687	\\
2433456.5518	&	1950.477	&	B	&	14.6	&	HCO(A27259)	&	0.866	\\
2433456.5658	&	1950.477	&	B	&	14.5	&	HCO(A27260)	&	0.869	\\
2440408.6477	&	1969.512	&	B	&	14.73	&	Mumford	&	0.362	\\
2441741.313	&	1973.160	&	B	&	11.47	&	Vidal et al.	&	0.678	\\
2441763.208	&	1973.220	&	B	&	11.01	&	Vidal et al.	&	0.462	\\
2441813.096	&	1973.357	&	B	&	11.61	&	Vidal et al.	&	0.083	\\
2441831.063	&	1973.406	&	B	&	13.0	&	Vidal et al.	&	0.189	\\
2441852.063	&	1973.464	&	B	&	14.30	&	Vidal et al.	&	0.818	\\
2441856.146	&	1973.475	&	B	&	14.30	&	Vidal et al.	&	0.523	\\
2442591.8236	&	1975.489	&	B	&	14.57	&	Landolt	&	0.666	\\
2442591.8257	&	1975.489	&	B	&	14.53	&	Landolt	&	0.666	\\
2442592.8219	&	1975.492	&	B	&	14.67	&	Landolt	&	0.838	\\
2442592.8238	&	1975.492	&	B	&	14.75	&	Landolt	&	0.838	\\
2442904.5	&	1976.345	&	B	&	14.94	&	Walker	&	0.703	\\
2443303.9012	&	1977.438	&	B	&	14.58	&	Landolt	&	0.729	\\
2443303.9030	&	1977.438	&	B	&	14.57	&	Landolt	&	0.729	\\
2443303.9063	&	1977.438	&	B	&	14.49	&	Landolt	&	0.730	\\
2443303.9080	&	1977.438	&	B	&	14.61	&	Landolt	&	0.730	\\
2443305.8919	&	1977.444	&	B	&	14.84	&	Landolt	&	0.073	\\
2443305.8937	&	1977.444	&	B	&	14.76	&	Landolt	&	0.073	\\
2443305.8978	&	1977.444	&	B	&	14.76	&	Landolt	&	0.074	\\
2443305.9004	&	1977.444	&	B	&	14.70	&	Landolt	&	0.074	\\
2443607.8800	&	1978.271	&	B	&	14.38	&	Landolt	&	0.263	\\
2443607.9054	&	1978.271	&	B	&	14.45	&	Landolt	&	0.268	\\
2443608.9041	&	1978.274	&	B	&	14.35	&	Landolt	&	0.440	\\
2443671.9011	&	1978.446	&	B	&	14.70	&	Landolt	&	0.328	\\
2443814.5101	&	1978.836	&	B	&	14.55	&	Landolt	&	0.974	\\
2444048.7141	&	1979.478	&	B	&	14.46	&	Landolt	&	0.449	\\
2444048.7202	&	1979.478	&	B	&	14.51	&	Landolt	&	0.450	\\
2444049.6943	&	1979.480	&	B	&	14.59	&	Landolt	&	0.619	\\
2444049.7006	&	1979.480	&	B	&	14.46	&	Landolt	&	0.620	\\
2444428.7282	&	1980.518	&	B	&	14.86	&	Landolt	&	0.124	\\
2444494.6320	&	1980.699	&	B	&	14.74	&	Landolt	&	0.514	\\
2444494.6393	&	1980.699	&	B	&	14.74	&	Landolt	&	0.515	\\
2444498.5282	&	1980.709	&	B	&	14.60	&	Landolt	&	0.187	\\
2444498.5361	&	1980.709	&	B	&	14.54	&	Landolt	&	0.189	\\
2444765.8678	&	1981.441	&	B	&	14.78	&	Landolt	&	0.390	\\
2444826.6562	&	1981.608	&	B	&	14.11	&	Landolt	&	0.895	\\
2444826.6584	&	1981.608	&	B	&	14.57	&	Landolt	&	0.896	\\
2444903.5386	&	1981.818	&	B	&	14.61	&	Landolt	&	0.182	\\
2444905.5506	&	1981.824	&	B	&	14.77	&	Landolt	&	0.530	\\
2445226.5191	&	1982.702	&	B	&	15.06	&	Landolt	&	0.000	\\
2445520.7701	&	1983.508	&	B	&	14.78	&	Landolt	&	0.854	\\
2445597.5831	&	1983.718	&	B	&	14.90	&	Landolt	&	0.129	\\
2445628.5530	&	1983.803	&	B	&	14.76	&	Landolt	&	0.481	\\
2445835.8185	&	1984.370	&	B	&	14.70	&	Landolt	&	0.301	\\
2445835.8293	&	1984.371	&	B	&	14.63	&	Landolt	&	0.303	\\
		\hline
	\end{tabular}
\end{table}

\begin{table}
	\centering
	\contcaption{Magnitudes}
	\label{tab:continued}
	\begin{tabular}{llllll} 
		\hline
		JD & Year & B. & Mag & Source & Phase\\
		\hline

2445835.8348	&	1984.371	&	B	&	14.67	&	Landolt	&	0.304	\\
2445835.8435	&	1984.371	&	B	&	14.69	&	Landolt	&	0.305	\\
2445978.5803	&	1984.761	&	B	&	14.98	&	Landolt	&	0.973	\\
2445984.5417	&	1984.778	&	B	&	15.05	&	Landolt	&	0.004	\\
2446574.8466	&	1986.394	&	B	&	15.19	&	Landolt	&	0.021	\\
2447457.5659	&	1988.811	&	B	&	15.02	&	Landolt	&	0.574	\\
2448055.8171	&	1990.449	&	B	&	15.12	&	Landolt	&	0.965	\\
2449154.8703	&	1993.458	&	B	&	14.61	&	Landolt	&	0.904	\\
2449929.4977	&	1995.578	&	B	&	14.50	&	Landolt	&	0.776	\\
2449929.5044	&	1995.578	&	B	&	14.50	&	Landolt	&	0.777	\\
2450316.5170	&	1996.638	&	B	&	14.56	&	Landolt	&	0.661	\\
2450316.5288	&	1996.638	&	B	&	14.55	&	Landolt	&	0.663	\\
2451081.5326	&	1998.732	&	B	&	14.79	&	Landolt	&	0.871	\\
2452189.5632	&	2001.766	&	B	&	14.75	&	Landolt	&	0.361	\\
2453600.7049	&	2005.630	&	B	&	14.71	&	SMARTS	&	0.234	\\
2453603.6424	&	2005.638	&	B	&	14.71	&	SMARTS	&	0.742	\\
2453634.6350	&	2005.722	&	B	&	14.89	&	SMARTS	&	0.098	\\
2453799.8943	&	2006.175	&	B	&	14.79	&	SMARTS	&	0.658	\\
2440408.6477	&	1969.512	&	V	&	13.64	&	Mumford	&	0.362	\\
2441741.313	&	1973.160	&	V	&	11.25	&	Vidal et al.	&	0.678	\\
2441757.8746	&	1973.206	&	V	&	10.72	&	Landolt	&	0.540	\\
2441760.8877	&	1973.214	&	V	&	10.55	&	Landolt	&	0.061	\\
2441761.8796	&	1973.217	&	V	&	10.77	&	Landolt	&	0.232	\\
2441763.208	&	1973.220	&	V	&	10.70	&	Vidal et al.	&	0.462	\\
2441764.8920	&	1973.225	&	V	&	10.56	&	Landolt	&	0.753	\\
2441767.8645	&	1973.233	&	V	&	10.51	&	Landolt	&	0.266	\\
2441771.8852	&	1973.244	&	V	&	10.36	&	Landolt	&	0.961	\\
2441813.096	&	1973.357	&	V	&	11.20	&	Vidal et al.	&	0.083	\\
2441831.063	&	1973.406	&	V	&	12.3	&	Vidal et al.	&	0.189	\\
2441852.063	&	1973.464	&	V	&	13.40	&	Vidal et al.	&	0.818	\\
2441856.146	&	1973.475	&	V	&	13.60	&	Vidal et al.	&	0.523	\\
2442141.8888	&	1974.257	&	V	&	13.84	&	Landolt	&	0.906	\\
2442141.8980	&	1974.257	&	V	&	13.52	&	Landolt	&	0.908	\\
2442151.8821	&	1974.284	&	V	&	13.33	&	Landolt	&	0.633	\\
2442152.9036	&	1974.287	&	V	&	13.75	&	Landolt	&	0.810	\\
2442591.8236	&	1975.489	&	V	&	13.50	&	Landolt	&	0.666	\\
2442591.8257	&	1975.489	&	V	&	13.46	&	Landolt	&	0.666	\\
2442592.8219	&	1975.492	&	V	&	13.71	&	Landolt	&	0.838	\\
2442592.8238	&	1975.492	&	V	&	13.72	&	Landolt	&	0.838	\\
2442904.5	&	1976.345	&	V	&	14.1	&	Walker	&	0.703	\\
2443303.9012	&	1977.438	&	V	&	13.46	&	Landolt	&	0.729	\\
2443303.9030	&	1977.438	&	V	&	13.38	&	Landolt	&	0.729	\\
2443303.9063	&	1977.438	&	V	&	13.38	&	Landolt	&	0.730	\\
2443303.9080	&	1977.438	&	V	&	13.54	&	Landolt	&	0.730	\\
2443305.8919	&	1977.444	&	V	&	13.63	&	Landolt	&	0.073	\\
2443305.8937	&	1977.444	&	V	&	13.61	&	Landolt	&	0.073	\\
2443305.8978	&	1977.444	&	V	&	13.70	&	Landolt	&	0.074	\\
2443305.9004	&	1977.444	&	V	&	13.72	&	Landolt	&	0.074	\\
2443607.8800	&	1978.271	&	V	&	13.19	&	Landolt	&	0.263	\\
2443607.9054	&	1978.271	&	V	&	13.20	&	Landolt	&	0.268	\\
2443608.9041	&	1978.274	&	V	&	12.82	&	Landolt	&	0.440	\\
2443671.9011	&	1978.446	&	V	&	13.46	&	Landolt	&	0.328	\\
2443814.5101	&	1978.836	&	V	&	13.69	&	Landolt	&	0.974	\\
2444048.7141	&	1979.478	&	V	&	13.44	&	Landolt	&	0.449	\\
2444048.7202	&	1979.478	&	V	&	13.42	&	Landolt	&	0.450	\\
2444049.6943	&	1979.480	&	V	&	13.51	&	Landolt	&	0.619	\\
2444049.7006	&	1979.480	&	V	&	13.40	&	Landolt	&	0.620	\\
2444428.7282	&	1980.518	&	V	&	13.75	&	Landolt	&	0.124	\\
2444494.6320	&	1980.699	&	V	&	13.60	&	Landolt	&	0.514	\\
2444494.6393	&	1980.699	&	V	&	13.60	&	Landolt	&	0.515	\\
2444498.5282	&	1980.709	&	V	&	13.47	&	Landolt	&	0.187	\\
2444498.5361	&	1980.709	&	V	&	13.42	&	Landolt	&	0.189	\\
2444765.8678	&	1981.441	&	V	&	13.67	&	Landolt	&	0.390	\\
2444826.6562	&	1981.608	&	V	&	13.05	&	Landolt	&	0.895	\\
		\hline
	\end{tabular}
\end{table}

\begin{table}
	\centering
	\contcaption{Magnitudes}
	\label{tab:continued}
	\begin{tabular}{llllll} 
		\hline
		JD & Year & B. & Mag & Source & Phase\\
		\hline

2444826.6584	&	1981.608	&	V	&	13.39	&	Landolt	&	0.896	\\
2444903.5386	&	1981.818	&	V	&	13.53	&	Landolt	&	0.182	\\
2444905.5506	&	1981.824	&	V	&	13.65	&	Landolt	&	0.530	\\
2445226.5191	&	1982.702	&	V	&	14.00	&	Landolt	&	0.000	\\
2445201.3700	&	1982.633	&	V	&	14.17	&	Landolt	&	0.654	\\
2445289.2318	&	1982.874	&	V	&	14.38	&	Landolt	&	0.839	\\
2445377.0936	&	1983.115	&	V	&	14.60	&	Landolt	&	0.023	\\
2445835.8185	&	1984.370	&	V	&	13.63	&	Landolt	&	0.301	\\
2445835.8293	&	1984.371	&	V	&	13.65	&	Landolt	&	0.303	\\
2445835.8348	&	1984.371	&	V	&	13.63	&	Landolt	&	0.304	\\
2445835.8435	&	1984.371	&	V	&	13.61	&	Landolt	&	0.305	\\
2445978.5803	&	1984.761	&	V	&	13.69	&	Landolt	&	0.973	\\
2445984.5417	&	1984.778	&	V	&	13.91	&	Landolt	&	0.004	\\
2446574.8466	&	1986.394	&	V	&	13.96	&	Landolt	&	0.021	\\
2447457.5659	&	1988.811	&	V	&	13.82	&	Landolt	&	0.574	\\
2448055.8171	&	1990.449	&	V	&	13.96	&	Landolt	&	0.965	\\
2449154.8703	&	1993.458	&	V	&	13.65	&	Landolt	&	0.904	\\
2449929.4977	&	1995.578	&	V	&	13.45	&	Landolt	&	0.776	\\
2449929.5044	&	1995.578	&	V	&	13.40	&	Landolt	&	0.777	\\
2450316.5170	&	1996.638	&	V	&	13.45	&	Landolt	&	0.661	\\
2450316.5288	&	1996.638	&	V	&	13.46	&	Landolt	&	0.663	\\
2451081.5326	&	1998.732	&	V	&	13.68	&	Landolt	&	0.871	\\
2451434.5438	&	1999.699	&	V	&	13.80	&	AAVSO(OCN)	&	0.878	\\
2451648.8826	&	2000.286	&	V	&	13.70	&	AAVSO(OCN)	&	0.920	\\
2451664.8479	&	2000.329	&	V	&	13.60	&	AAVSO(OCN)	&	0.680	\\
2451691.8069	&	2000.403	&	V	&	13.60	&	AAVSO(OCN)	&	0.339	\\
2451713.7299	&	2000.463	&	V	&	13.70	&	AAVSO(OCN)	&	0.127	\\
2451723.6813	&	2000.491	&	V	&	13.70	&	AAVSO(OCN)	&	0.847	\\
2451737.6875	&	2000.529	&	V	&	13.70	&	AAVSO(OCN)	&	0.268	\\
2451778.0903	&	2000.640	&	V	&	13.50	&	AAVSO(LAQ)	&	0.250	\\
2452189.5632	&	2001.766	&	V	&	13.64	&	Landolt	&	0.361	\\
2453600.7065	&	2005.630	&	V	&	13.64	&	SMARTS	&	0.234	\\
2453603.6440	&	2005.638	&	V	&	13.63	&	SMARTS	&	0.742	\\
2453634.6367	&	2005.722	&	V	&	13.80	&	SMARTS	&	0.098	\\
2453799.8960	&	2006.175	&	V	&	13.70	&	SMARTS	&	0.658	\\
2455308.9231	&	2010.306	&	V	&	13.78	&	AAVSO(DKS)	&	0.447	\\
2455309.8986	&	2010.309	&	V	&	13.76	&	AAVSO(DKS)	&	0.615	\\
2455313.9209	&	2010.320	&	V	&	13.70	&	AAVSO(DKS)	&	0.310	\\
2455320.9031	&	2010.339	&	V	&	13.78	&	AAVSO(DKS)	&	0.517	\\
2455325.9127	&	2010.353	&	V	&	13.71	&	AAVSO(DKS)	&	0.383	\\
2455327.8169	&	2010.358	&	V	&	13.66	&	AAVSO(DKS)	&	0.712	\\
2455329.8231	&	2010.364	&	V	&	13.80	&	AAVSO(DKS)	&	0.059	\\
2455330.7890	&	2010.366	&	V	&	13.30	&	AAVSO(DKS)	&	0.225	\\
2455334.8308	&	2010.377	&	V	&	13.81	&	AAVSO(DKS)	&	0.924	\\
2455335.8125	&	2010.380	&	V	&	13.72	&	AAVSO(DKS)	&	0.094	\\
2455336.7712	&	2010.383	&	V	&	13.48	&	AAVSO(DKS)	&	0.259	\\
2455338.7857	&	2010.388	&	V	&	13.70	&	AAVSO(DKS)	&	0.607	\\
2455339.7801	&	2010.391	&	V	&	13.57	&	AAVSO(DKS)	&	0.779	\\
2455356.7551	&	2010.437	&	V	&	13.66	&	AAVSO(DKS)	&	0.713	\\
2455359.7321	&	2010.446	&	V	&	13.57	&	AAVSO(DKS)	&	0.227	\\
2455455.5512	&	2010.708	&	V	&	13.72	&	AAVSO(DKS)	&	0.787	\\
2455456.5495	&	2010.711	&	V	&	13.94	&	AAVSO(DKS)	&	0.959	\\
2455457.5195	&	2010.713	&	V	&	13.78	&	AAVSO(DKS)	&	0.127	\\
2455459.5444	&	2010.719	&	V	&	13.82	&	AAVSO(DKS)	&	0.477	\\
2455460.5502	&	2010.722	&	V	&	13.71	&	AAVSO(DKS)	&	0.651	\\
2455461.5172	&	2010.724	&	V	&	13.70	&	AAVSO(DKS)	&	0.818	\\
2455470.5160	&	2010.749	&	V	&	13.71	&	AAVSO(DKS)	&	0.373	\\
2455471.5286	&	2010.752	&	V	&	13.87	&	AAVSO(DKS)	&	0.548	\\
2455472.5281	&	2010.754	&	V	&	13.70	&	AAVSO(DKS)	&	0.721	\\
2455473.5100	&	2010.757	&	V	&	13.76	&	AAVSO(DKS)	&	0.890	\\
2455474.5338	&	2010.760	&	V	&	13.88	&	AAVSO(DKS)	&	0.067	\\
2455475.5618	&	2010.763	&	V	&	13.60	&	AAVSO(DKS)	&	0.245	\\
2455476.5461	&	2010.765	&	V	&	13.81	&	AAVSO(DKS)	&	0.415	\\
2455477.5144	&	2010.768	&	V	&	13.71	&	AAVSO(DKS)	&	0.582	\\
		\hline
	\end{tabular}
\end{table}

\begin{table}
	\centering
	\contcaption{Magnitudes}
	\label{tab:continued}
	\begin{tabular}{llllll} 
		\hline
		JD & Year & B. & Mag & Source & Phase\\
		\hline

2455479.5069	&	2010.773	&	V	&	13.85	&	AAVSO(DKS)	&	0.927	\\
2455480.5061	&	2010.776	&	V	&	13.75	&	AAVSO(DKS)	&	0.099	\\
2455484.5226	&	2010.787	&	V	&	13.66	&	AAVSO(DKS)	&	0.794	\\
2455487.5166	&	2010.795	&	V	&	13.64	&	AAVSO(DKS)	&	0.311	\\
2455490.5120	&	2010.804	&	V	&	13.64	&	AAVSO(DKS)	&	0.829	\\
2455493.5040	&	2010.812	&	V	&	13.70	&	AAVSO(DKS)	&	0.346	\\
2455497.5027	&	2010.823	&	V	&	13.91	&	AAVSO(DKS)	&	0.037	\\
2455665.9329	&	2011.284	&	V	&	13.61	&	AAVSO(DKS)	&	0.145	\\
2455683.9171	&	2011.333	&	V	&	13.57	&	AAVSO(DKS)	&	0.253	\\
2455711.7320	&	2011.409	&	V	&	13.90	&	AAVSO(DKS)	&	0.060	\\
2455839.5289	&	2011.759	&	V	&	13.64	&	AAVSO(DKS)	&	0.145	\\
2455848.4988	&	2011.784	&	V	&	13.68	&	AAVSO(DKS)	&	0.695	\\
2455849.5108	&	2011.786	&	V	&	13.68	&	AAVSO(DKS)	&	0.870	\\
2455855.5030	&	2011.803	&	V	&	13.80	&	AAVSO(DKS)	&	0.906	\\
2456007.9329	&	2012.220	&	V	&	13.67	&	AAVSO(DKS)	&	0.249	\\
2456008.9337	&	2012.223	&	V	&	13.80	&	AAVSO(DKS)	&	0.422	\\
2456013.9341	&	2012.237	&	V	&	13.93	&	AAVSO(DKS)	&	0.286	\\
2456014.9174	&	2012.239	&	V	&	14.00	&	AAVSO(DKS)	&	0.456	\\
2456015.9218	&	2012.242	&	V	&	13.90	&	AAVSO(DKS)	&	0.629	\\
2456021.9210	&	2012.259	&	V	&	14.00	&	AAVSO(DKS)	&	0.666	\\
2456026.9298	&	2012.272	&	V	&	13.97	&	AAVSO(DKS)	&	0.532	\\
2456032.9229	&	2012.289	&	V	&	13.93	&	AAVSO(DKS)	&	0.567	\\
2456033.8705	&	2012.291	&	V	&	13.87	&	AAVSO(DKS)	&	0.731	\\
2456034.8872	&	2012.294	&	V	&	13.83	&	AAVSO(DKS)	&	0.907	\\
2456045.8704	&	2012.324	&	V	&	13.72	&	AAVSO(DKS)	&	0.805	\\
2456048.8698	&	2012.332	&	V	&	13.68	&	AAVSO(DKS)	&	0.323	\\
2456049.8710	&	2012.335	&	V	&	13.83	&	AAVSO(DKS)	&	0.496	\\
2456050.8711	&	2012.338	&	V	&	13.60	&	AAVSO(DKS)	&	0.669	\\
2456053.8694	&	2012.346	&	V	&	13.70	&	AAVSO(DKS)	&	0.187	\\
2456054.8696	&	2012.349	&	V	&	13.90	&	AAVSO(DKS)	&	0.360	\\
2456065.7818	&	2012.379	&	V	&	13.50	&	AAVSO(DKS)	&	0.246	\\
2456067.7816	&	2012.384	&	V	&	13.70	&	AAVSO(DKS)	&	0.592	\\
2456069.8426	&	2012.390	&	V	&	13.61	&	AAVSO(DKS)	&	0.948	\\
2456070.8146	&	2012.392	&	V	&	13.69	&	AAVSO(DKS)	&	0.116	\\
2456071.8113	&	2012.395	&	V	&	13.80	&	AAVSO(DKS)	&	0.288	\\
2456074.8192	&	2012.403	&	V	&	13.30	&	AAVSO(DKS)	&	0.808	\\
2456081.8211	&	2012.423	&	V	&	14.20	&	AAVSO(DKS)	&	0.018	\\
2456082.8294	&	2012.425	&	V	&	13.84	&	AAVSO(DKS)	&	0.192	\\
2456096.7593	&	2012.463	&	V	&	13.97	&	AAVSO(DKS)	&	0.599	\\
2456097.7575	&	2012.466	&	V	&	13.75	&	AAVSO(DKS)	&	0.772	\\
2456106.7569	&	2012.491	&	V	&	13.87	&	AAVSO(DKS)	&	0.327	\\
2456107.7557	&	2012.494	&	V	&	13.93	&	AAVSO(DKS)	&	0.500	\\
2456116.7577	&	2012.518	&	V	&	13.93	&	AAVSO(DKS)	&	0.056	\\
2456120.7582	&	2012.529	&	V	&	13.61	&	AAVSO(DKS)	&	0.747	\\
2456365.9450	&	2013.200	&	V	&	13.74	&	AAVSO(DKS)	&	0.120	\\
2456366.9468	&	2013.203	&	V	&	13.76	&	AAVSO(DKS)	&	0.293	\\
2456367.9470	&	2013.206	&	V	&	13.50	&	AAVSO(DKS)	&	0.466	\\
2456368.9487	&	2013.209	&	V	&	13.63	&	AAVSO(DKS)	&	0.639	\\
2456380.9454	&	2013.241	&	V	&	13.53	&	AAVSO(DKS)	&	0.712	\\
2456381.9489	&	2013.244	&	V	&	13.91	&	AAVSO(DKS)	&	0.886	\\
2456398.9269	&	2013.291	&	V	&	13.56	&	AAVSO(DKS)	&	0.820	\\
2456419.9142	&	2013.348	&	V	&	13.55	&	AAVSO(DKS)	&	0.447	\\
2456420.9136	&	2013.351	&	V	&	13.54	&	AAVSO(DKS)	&	0.619	\\
2456421.9141	&	2013.354	&	V	&	13.54	&	AAVSO(DKS)	&	0.792	\\
2456422.7946	&	2013.356	&	V	&	13.63	&	AAVSO(DKS)	&	0.944	\\
2456427.8490	&	2013.370	&	V	&	13.52	&	AAVSO(DKS)	&	0.818	\\
2456428.8892	&	2013.373	&	V	&	13.81	&	AAVSO(DKS)	&	0.998	\\
2456430.8962	&	2013.378	&	V	&	13.64	&	AAVSO(DKS)	&	0.344	\\
2456436.8531	&	2013.395	&	V	&	13.71	&	AAVSO(DKS)	&	0.374	\\
2456444.8873	&	2013.417	&	V	&	13.44	&	AAVSO(DKS)	&	0.762	\\
2456457.7355	&	2013.452	&	V	&	13.81	&	AAVSO(DKS)	&	0.983	\\
2456545.5169	&	2013.692	&	V	&	13.68	&	AAVSO(DKS)	&	0.153	\\
2456577.4927	&	2013.780	&	V	&	13.61	&	AAVSO(DKS)	&	0.679	\\
2456584.4894	&	2013.799	&	V	&	13.76	&	AAVSO(DKS)	&	0.888	\\
		\hline
	\end{tabular}
\end{table}

\begin{table}
	\centering
	\contcaption{Magnitudes}
	\label{tab:continued}
	\begin{tabular}{llllll} 
		\hline
		JD & Year & B. & Mag & Source & Phase\\
		\hline

2456593.4745	&	2013.823	&	V	&	13.57	&	AAVSO(DKS)	&	0.441	\\
2456600.4732	&	2013.843	&	V	&	13.63	&	AAVSO(DKS)	&	0.651	\\
2456725.9176	&	2014.186	&	V	&	13.30	&	AAVSO(DKS)	&	0.330	\\
2456727.9478	&	2014.192	&	V	&	13.57	&	AAVSO(DKS)	&	0.680	\\
2456730.9446	&	2014.200	&	V	&	13.72	&	AAVSO(DKS)	&	0.198	\\
2456731.9467	&	2014.202	&	V	&	13.79	&	AAVSO(DKS)	&	0.372	\\
2456737.9480	&	2014.219	&	V	&	13.75	&	AAVSO(DKS)	&	0.409	\\
2456742.9417	&	2014.233	&	V	&	13.68	&	AAVSO(DKS)	&	0.272	\\
2456748.9312	&	2014.249	&	V	&	13.58	&	AAVSO(DKS)	&	0.307	\\
2456749.9431	&	2014.252	&	V	&	13.76	&	AAVSO(DKS)	&	0.482	\\
2456750.9397	&	2014.254	&	V	&	13.59	&	AAVSO(DKS)	&	0.654	\\
2456752.9381	&	2014.260	&	V	&	13.88	&	AAVSO(DKS)	&	0.999	\\
2456764.8898	&	2014.293	&	V	&	13.70	&	AAVSO(DKS)	&	0.065	\\
2456782.9033	&	2014.342	&	V	&	13.38	&	AAVSO(DKS)	&	0.178	\\
2456783.8438	&	2014.345	&	V	&	13.57	&	AAVSO(DKS)	&	0.340	\\
2456796.7698	&	2014.380	&	V	&	14.00	&	AAVSO(DKS)	&	0.574	\\
2456797.8972	&	2014.383	&	V	&	13.80	&	AAVSO(DKS)	&	0.769	\\
2456798.8945	&	2014.386	&	V	&	13.71	&	AAVSO(DKS)	&	0.941	\\
2456799.8467	&	2014.388	&	V	&	13.66	&	AAVSO(DKS)	&	0.106	\\
2456800.8993	&	2014.391	&	V	&	13.90	&	AAVSO(DKS)	&	0.288	\\
2456806.7720	&	2014.407	&	V	&	13.62	&	AAVSO(DKS)	&	0.303	\\
2456941.5016	&	2014.776	&	V	&	13.69	&	AAVSO(DKS)	&	0.586	\\
2456942.4923	&	2014.779	&	V	&	13.31	&	AAVSO(DKS)	&	0.758	\\
2457090.9524	&	2015.185	&	V	&	13.60	&	AAVSO(DKS)	&	0.414	\\
2457102.9476	&	2015.218	&	V	&	13.83	&	AAVSO(DKS)	&	0.487	\\
2457110.9374	&	2015.240	&	V	&	13.77	&	AAVSO(DKS)	&	0.868	\\
2457112.9131	&	2015.245	&	V	&	13.60	&	AAVSO(DKS)	&	0.210	\\
2457113.8931	&	2015.248	&	V	&	13.73	&	AAVSO(DKS)	&	0.379	\\
2457114.8972	&	2015.251	&	V	&	13.93	&	AAVSO(DKS)	&	0.552	\\
2457116.8901	&	2015.256	&	V	&	13.85	&	AAVSO(DKS)	&	0.897	\\
2457121.9261	&	2015.270	&	V	&	13.84	&	AAVSO(DKS)	&	0.767	\\
2457126.9227	&	2015.284	&	V	&	13.76	&	AAVSO(DKS)	&	0.631	\\
2457137.9125	&	2015.314	&	V	&	13.78	&	AAVSO(DKS)	&	0.530	\\
2457145.8918	&	2015.336	&	V	&	13.90	&	AAVSO(DKS)	&	0.909	\\
2457150.9075	&	2015.350	&	V	&	13.65	&	AAVSO(DKS)	&	0.776	\\
2457151.9056	&	2015.352	&	V	&	13.94	&	AAVSO(DKS)	&	0.948	\\
2457156.8188	&	2015.366	&	V	&	13.74	&	AAVSO(DKS)	&	0.797	\\
2457157.8475	&	2015.369	&	V	&	14.06	&	AAVSO(DKS)	&	0.975	\\
2457160.8442	&	2015.377	&	V	&	13.82	&	AAVSO(DKS)	&	0.493	\\
2457161.9065	&	2015.380	&	V	&	13.58	&	AAVSO(DKS)	&	0.676	\\
2457164.7754	&	2015.387	&	V	&	13.62	&	AAVSO(DKS)	&	0.172	\\
2457166.9078	&	2015.393	&	V	&	13.80	&	AAVSO(DKS)	&	0.541	\\
2457167.9015	&	2015.396	&	V	&	13.64	&	AAVSO(DKS)	&	0.713	\\
2457170.8751	&	2015.404	&	V	&	13.63	&	AAVSO(DKS)	&	0.226	\\
2457261.66	&	2015.653	&	V	&	13.73	&	HRPO	&	0.916	\\
2457269.61	&	2015.674	&	V	&	13.32	&	HRPO	&	0.290	\\
2457274.62	&	2015.688	&	V	&	13.55	&	HRPO	&	0.156	\\
2457280.58	&	2015.705	&	V	&	13.76	&	HRPO	&	0.186	\\
2457283.59	&	2015.713	&	V	&	13.64	&	HRPO	&	0.706	\\
2457288.55	&	2015.726	&	V	&	13.78	&	HRPO	&	0.562	\\
2457297.53	&	2015.751	&	V	&	13.69	&	HRPO	&	0.114	\\
2457309.52	&	2015.784	&	V	&	13.58	&	HRPO	&	0.187	\\
2457311.4825	&	2015.789	&	V	&	13.83	&	AAVSO(DKS)	&	0.526	\\
2457311.54	&	2015.789	&	V	&	13.76	&	HRPO	&	0.535	\\
2457313.4823	&	2015.795	&	V	&	13.68	&	AAVSO(DKS)	&	0.872	\\
2457316.51	&	2015.803	&	V	&	13.69	&	HRPO	&	0.395	\\
2441757.8746	&	1973.206	&	R	&	10.36	&	Landolt	&	0.540	\\
2441760.8877	&	1973.214	&	R	&	10.11	&	Landolt	&	0.061	\\
2441761.8796	&	1973.217	&	R	&	10.15	&	Landolt	&	0.232	\\
2441764.8920	&	1973.225	&	R	&	10.14	&	Landolt	&	0.753	\\
2441767.8645	&	1973.233	&	R	&	10.08	&	Landolt	&	0.266	\\
2441771.8852	&	1973.244	&	R	&	9.94	&	Landolt	&	0.961	\\
2442141.8888	&	1974.257	&	R	&	12.63	&	Landolt	&	0.906	\\
2442141.8980	&	1974.257	&	R	&	12.53	&	Landolt	&	0.908	\\
		\hline
	\end{tabular}
\end{table}

\begin{table}
	\centering
	\contcaption{Magnitudes}
	\label{tab:continued}
	\begin{tabular}{llllll} 
		\hline
		JD & Year & B. & Mag & Source & Phase\\
		\hline

2442151.8821	&	1974.284	&	R	&	12.50	&	Landolt	&	0.633	\\
2442152.9036	&	1974.287	&	R	&	12.37	&	Landolt	&	0.810	\\
2443607.8800	&	1978.271	&	R	&	12.55	&	Landolt	&	0.263	\\
2443607.9054	&	1978.271	&	R	&	12.54	&	Landolt	&	0.268	\\
2443608.9041	&	1978.274	&	R	&	12.35	&	Landolt	&	0.440	\\
2443671.9011	&	1978.446	&	R	&	12.77	&	Landolt	&	0.328	\\
2443814.5101	&	1978.836	&	R	&	12.96	&	Landolt	&	0.974	\\
2444048.7141	&	1979.478	&	R	&	12.74	&	Landolt	&	0.449	\\
2444048.7202	&	1979.478	&	R	&	12.76	&	Landolt	&	0.450	\\
2444049.6943	&	1979.480	&	R	&	12.81	&	Landolt	&	0.619	\\
2444049.7006	&	1979.480	&	R	&	12.70	&	Landolt	&	0.620	\\
2444498.5282	&	1980.709	&	R	&	12.74	&	Landolt	&	0.187	\\
2444498.5361	&	1980.709	&	R	&	12.72	&	Landolt	&	0.189	\\
2444765.8678	&	1981.441	&	R	&	12.99	&	Landolt	&	0.390	\\
2444905.5506	&	1981.824	&	R	&	12.94	&	Landolt	&	0.530	\\
2445226.5191	&	1982.702	&	R	&	13.28	&	Landolt	&	0.000	\\
2445520.7701	&	1983.508	&	R	&	13.00	&	Landolt	&	0.854	\\
2445597.5831	&	1983.718	&	R	&	13.06	&	Landolt	&	0.129	\\
2445628.5530	&	1983.803	&	R	&	12.95	&	Landolt	&	0.481	\\
2445835.8185	&	1984.370	&	R	&	12.94	&	Landolt	&	0.301	\\
2445835.8293	&	1984.371	&	R	&	12.96	&	Landolt	&	0.303	\\
2445835.8348	&	1984.371	&	R	&	12.94	&	Landolt	&	0.304	\\
2445835.8435	&	1984.371	&	R	&	12.95	&	Landolt	&	0.305	\\
2445978.5803	&	1984.761	&	R	&	13.05	&	Landolt	&	0.973	\\
2445984.5417	&	1984.778	&	R	&	13.16	&	Landolt	&	0.004	\\
2446574.8466	&	1986.394	&	R	&	13.16	&	Landolt	&	0.021	\\
2447457.5659	&	1988.811	&	R	&	13.10	&	Landolt	&	0.574	\\
2448055.8171	&	1990.449	&	R	&	13.27	&	Landolt	&	0.965	\\
2449154.8703	&	1993.458	&	R	&	12.97	&	Landolt	&	0.904	\\
2450316.5170	&	1996.638	&	R	&	12.75	&	Landolt	&	0.661	\\
2450316.5288	&	1996.638	&	R	&	12.74	&	Landolt	&	0.663	\\
2451081.5326	&	1998.732	&	R	&	12.99	&	Landolt	&	0.871	\\
2452189.5632	&	2001.766	&	R	&	12.95	&	Landolt	&	0.361	\\
2453600.7078	&	2005.630	&	R	&	12.93	&	SMARTS	&	0.234	\\
2453603.6453	&	2005.638	&	R	&	12.93	&	SMARTS	&	0.742	\\
2453634.6379	&	2005.722	&	R	&	13.09	&	SMARTS	&	0.098	\\
2453799.8973	&	2006.175	&	R	&	13.01	&	SMARTS	&	0.658	\\
2453815.8725	&	2006.219	&	R	&	13.10	&	SMARTS	&	0.419	\\
2453815.8734	&	2006.219	&	R	&	13.11	&	SMARTS	&	0.419	\\
2453819.8201	&	2006.229	&	R	&	13.10	&	SMARTS	&	0.101	\\
2453819.8210	&	2006.229	&	R	&	13.10	&	SMARTS	&	0.101	\\
2453823.8685	&	2006.241	&	R	&	12.98	&	SMARTS	&	0.801	\\
2453823.8695	&	2006.241	&	R	&	12.98	&	SMARTS	&	0.801	\\
2453829.8324	&	2006.257	&	R	&	12.99	&	SMARTS	&	0.832	\\
2453829.8333	&	2006.257	&	R	&	12.99	&	SMARTS	&	0.832	\\
2453833.8293	&	2006.268	&	R	&	13.07	&	SMARTS	&	0.522	\\
2453833.8302	&	2006.268	&	R	&	13.08	&	SMARTS	&	0.522	\\
2453836.7916	&	2006.276	&	R	&	13.00	&	SMARTS	&	0.034	\\
2453836.7925	&	2006.276	&	R	&	12.99	&	SMARTS	&	0.034	\\
2453847.8550	&	2006.306	&	R	&	12.99	&	SMARTS	&	0.946	\\
2453847.8559	&	2006.306	&	R	&	12.99	&	SMARTS	&	0.946	\\
2453864.8312	&	2006.353	&	R	&	13.08	&	SMARTS	&	0.880	\\
2453864.8321	&	2006.353	&	R	&	13.07	&	SMARTS	&	0.880	\\
2453869.6997	&	2006.366	&	R	&	12.93	&	SMARTS	&	0.721	\\
2453869.7007	&	2006.366	&	R	&	12.93	&	SMARTS	&	0.722	\\
2453869.7018	&	2006.366	&	R	&	12.94	&	SMARTS	&	0.722	\\
2453869.7027	&	2006.366	&	R	&	12.94	&	SMARTS	&	0.722	\\
2453891.7340	&	2006.426	&	R	&	13.07	&	SMARTS	&	0.529	\\
2453891.7349	&	2006.426	&	R	&	13.07	&	SMARTS	&	0.529	\\
2453904.8303	&	2006.462	&	R	&	12.94	&	SMARTS	&	0.793	\\
2453904.8312	&	2006.462	&	R	&	12.94	&	SMARTS	&	0.793	\\
2441757.8746	&	1973.206	&	I	&	10.10	&	Landolt	&	0.540	\\
2441760.8877	&	1973.214	&	I	&	9.86	&	Landolt	&	0.061	\\
2441761.8796	&	1973.217	&	I	&	9.80	&	Landolt	&	0.232	\\
		\hline
	\end{tabular}
\end{table}

\begin{table}
	\centering
	\contcaption{Magnitudes}
	\label{tab:continued}
	\begin{tabular}{llllll} 
		\hline
		JD & Year & B. & Mag & Source & Phase\\
		\hline
2441764.8920	&	1973.225	&	I	&	9.78	&	Landolt	&	0.753	\\
2441767.8645	&	1973.233	&	I	&	9.75	&	Landolt	&	0.266	\\
2441771.8852	&	1973.244	&	I	&	9.70	&	Landolt	&	0.961	\\
2442141.8888	&	1974.257	&	I	&	11.64	&	Landolt	&	0.906	\\
2442141.8980	&	1974.257	&	I	&	11.70	&	Landolt	&	0.908	\\
2442151.8821	&	1974.284	&	I	&	11.84	&	Landolt	&	0.633	\\
2442152.9036	&	1974.287	&	I	&	11.56	&	Landolt	&	0.810	\\
2453600.7088	&	2005.630	&	I	&	12.32	&	SMARTS	&	0.235	\\
2453603.6464	&	2005.638	&	I	&	12.32	&	SMARTS	&	0.742	\\
2453634.6390	&	2005.722	&	I	&	12.45	&	SMARTS	&	0.098	\\
2453799.8984	&	2006.175	&	I	&	12.37	&	SMARTS	&	0.658	\\
2448491.08	&	1991.640	&	J	&	11.25	&	Harrison et al.	&	0.187	\\
2448491.08	&	1991.640	&	H	&	10.67	&	Harrison et al.	&	0.187	\\
2448491.08	&	1991.640	&	K	&	10.5	&	Harrison et al.	&	0.187	\\
2448491.08	&	1991.640	&	L	&	10.39	&	Harrison et al.	&	0.187	\\
		\hline
	\end{tabular}
\end{table}

\section{Light Curve Analysis}
The light curve of the 1919 event is shown in Figure 2, all as based on the HCO plates with correction of the magnitudes for the unresolved light of CD -29\degr 15053.  We see a normal nova light curve, apparently of the S-class with a smooth decline (Strope et al. 2008).  We see a linear decline at a rate of 9.3 magnitudes per year which is certainly greatly larger in amplitude and longer in duration than any DN event.  Furthermore, this is not the light curve of any known symbiotic nova event.  So this is the proof that V1017 Sgr had a CN eruption in 1919.

The amplitude is at least from B=6.43 at the brightest observed level to B=15.09 for the pre-eruption interval, for an amplitude of $>$8.66 mag.  The amplitude could be greatly larger if the true peak was substantially earlier than the first plate in 1919 (on 11 March 1919).  This is quite possible, and even likely, because V1017 Sgr was just coming out of its yearly close conjunction with the Sun, so the nova could easily have been greatly brighter with a peak anytime from November to February.  Indeed, for S-class novae with a linear decline over 200 days, essentially all have the start of the linear decline occurring more than 40 days after peak, and something like 2 mags below peak (Strope et al. 2010 Figure 3).  That is, the lack of a fast decline apparent in the V1017 Sgr light curve forces the real peak to be something like 2 mags brighter than B=6.43.  The next earliest useful plate is from 1 August 1918, before the star went behind the Sun, and provides no useful constraint.  Still, a simple extrapolation back puts it to second magnitude in October 1918, at which time it should have been picked up by many sky watchers with Sagittarius in the evening sky.  So the peak was at magnitude 6.4 to 2.0 or so, with an amplitude from roughly 8.7 to 13 mag or so.

\begin{figure}
	\includegraphics[width=1.1\columnwidth]{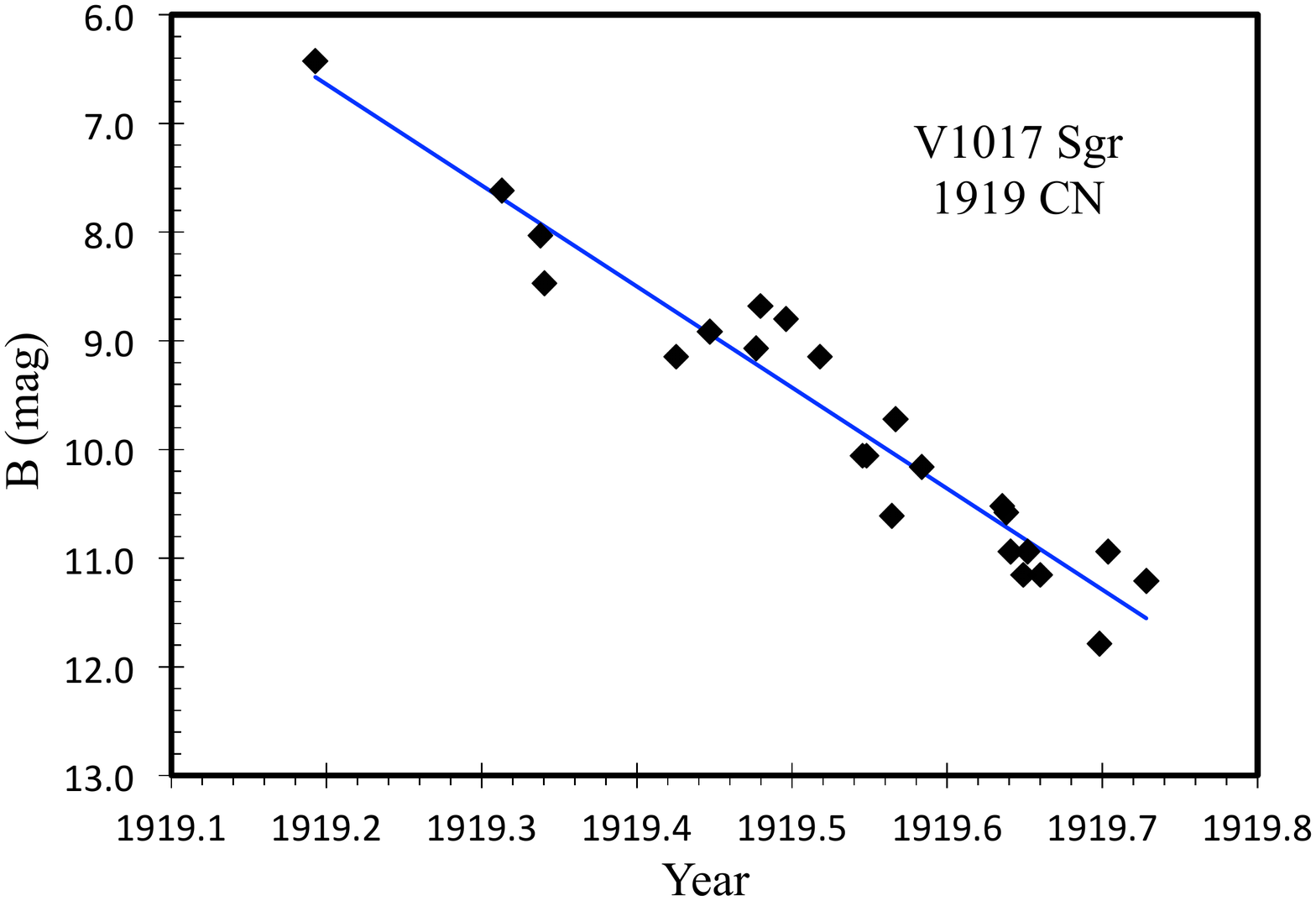}
    \caption{Classical nova eruption in 1919.  This B-band light curve is constructed from the HCO archival plates, and the magnitudes have been corrected for the unresolved star CD -29\degr 15053.  The only earlier and later light curve points are from the previous and next years, so they provide no useful constraint on peak time or magnitude.  We see an ordinary nova light curve, apparently in the 'smooth' S-class.  The main point of this light curve is to show that the 1919 even is not any sort of DN event, but is really a classical nova, so that V1017 Sgr certainly displays {\it both} CN and DN events.}
    \label{fig:Fig. 2}
\end{figure}

The light curves for the three known DN events (in 1901, 1973, and 1991) are shown in Figure 3.  All three events have identical light curve shapes, although the uncertainties are substantial.  The peak magnitude is near 10.4 mag, with an amplitude of near 3.3 mag, in both the visual and B bands.  The total duration is near 0.5 years.

This DN light curve (roughly triangular shape, several month duration, $\sim$3 mag amplitude) is the same as for GK Per.  GK Per is the only other CN that shows DN events.  Along with V1017 Sgr and X Ser, GK Per is the only CN with a known orbital period longer than 1.0 days.  This implies that both V1017 Sgr and GK Per both have the high accretion rate of CN plus DN events due to their long orbital periods.

Until recently, V1017 Sgr was the only CN that has observed DN events {\it before} the CN eruption.  (Other CN have been weakly claimed to have prior DN eruptions, but the cases for all of them have gone away; see Collazzi et al. 2009.) However, Mroz et al. (2016) have discovered that the CN V1213 Cen has many ordinary DN eruptions before the nova event. This has relevance for understanding CV evolution.  The essence is that DN are systems where the white dwarf is accumulating material from its companion star, so sooner or later the system must have a nova eruption.  In general, DN have a relatively low accretion rate while CN have a relatively high accretion rate.  For ordinary CVs, this means that the systems with low enough accretion rate to have DN eruptions must also have very long time scales between CN eruptions, so there will be virtually no systems discovered with DN events {\it before} the CN eruption.  (We can imagine that conditions {\it after} the eruption might have a higher accretion rate due to irradiation effects, so the presence of a DN event {\it after} the CN eruption won't tell us about the long-term accretion state of the CV.)  But V1017 Sgr is different from almost all CVs because it has a giant or sub-giant companion.  This forces the system to have a very large Roche lobe, so the accretion disk must also be very large, and so the outer portions of the accretion disk are cool enough to have a DN accretion instability despite a relatively high accretion rate.  With a giant companion, a CV system can have both DN and CN events.

\begin{figure}
	\includegraphics[width=1.0\columnwidth]{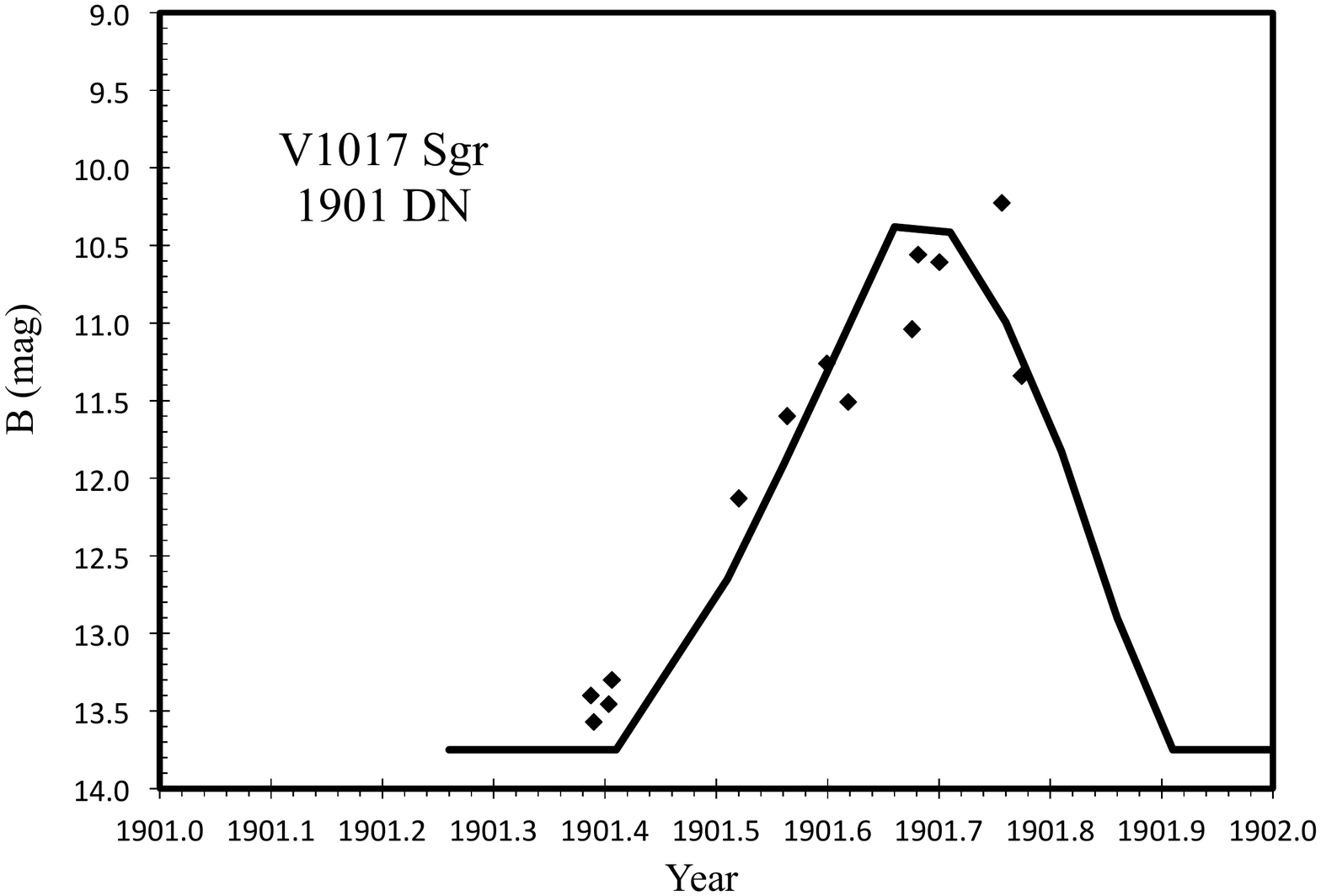}
	\includegraphics[width=1.0\columnwidth]{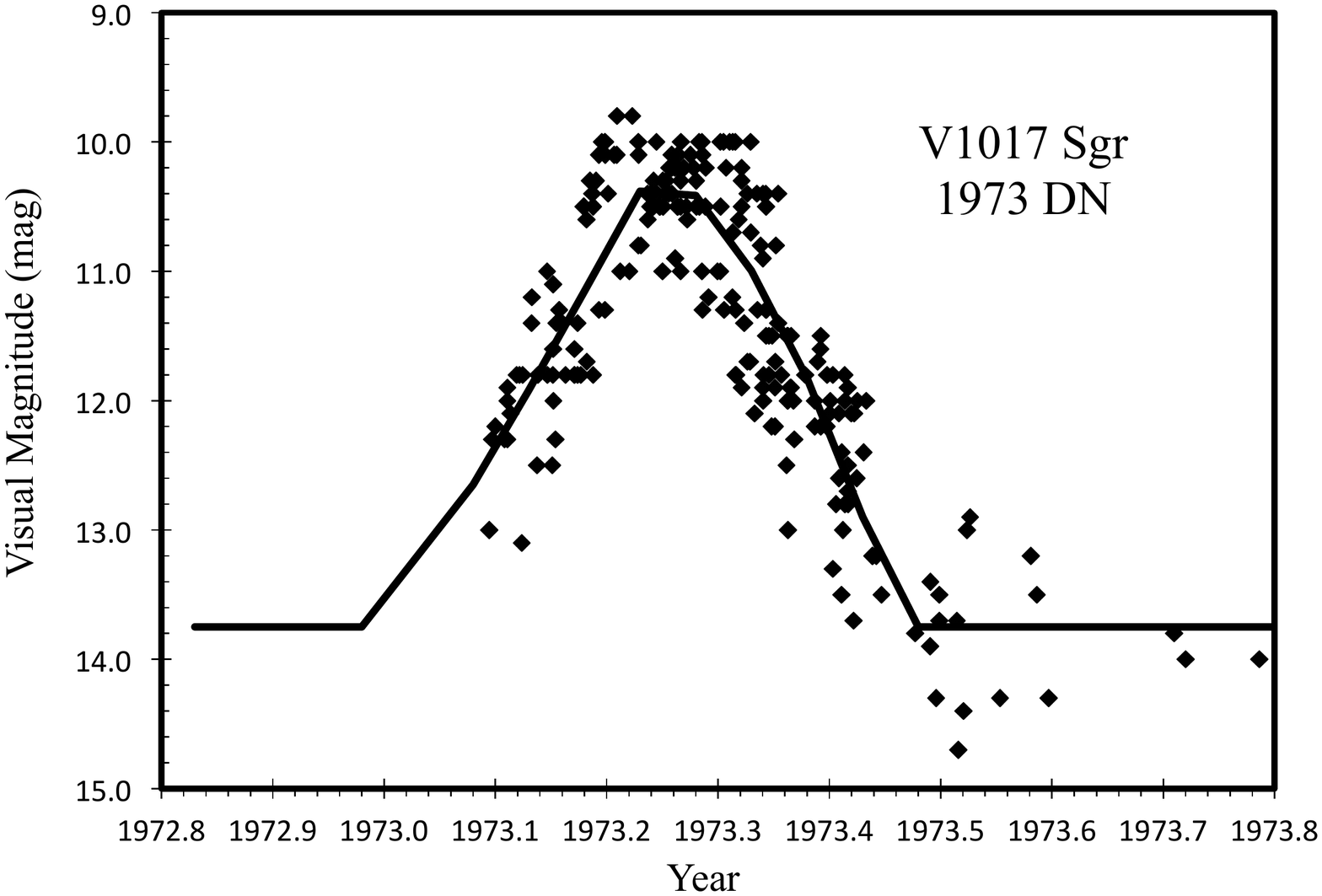}
	\includegraphics[width=1.0\columnwidth]{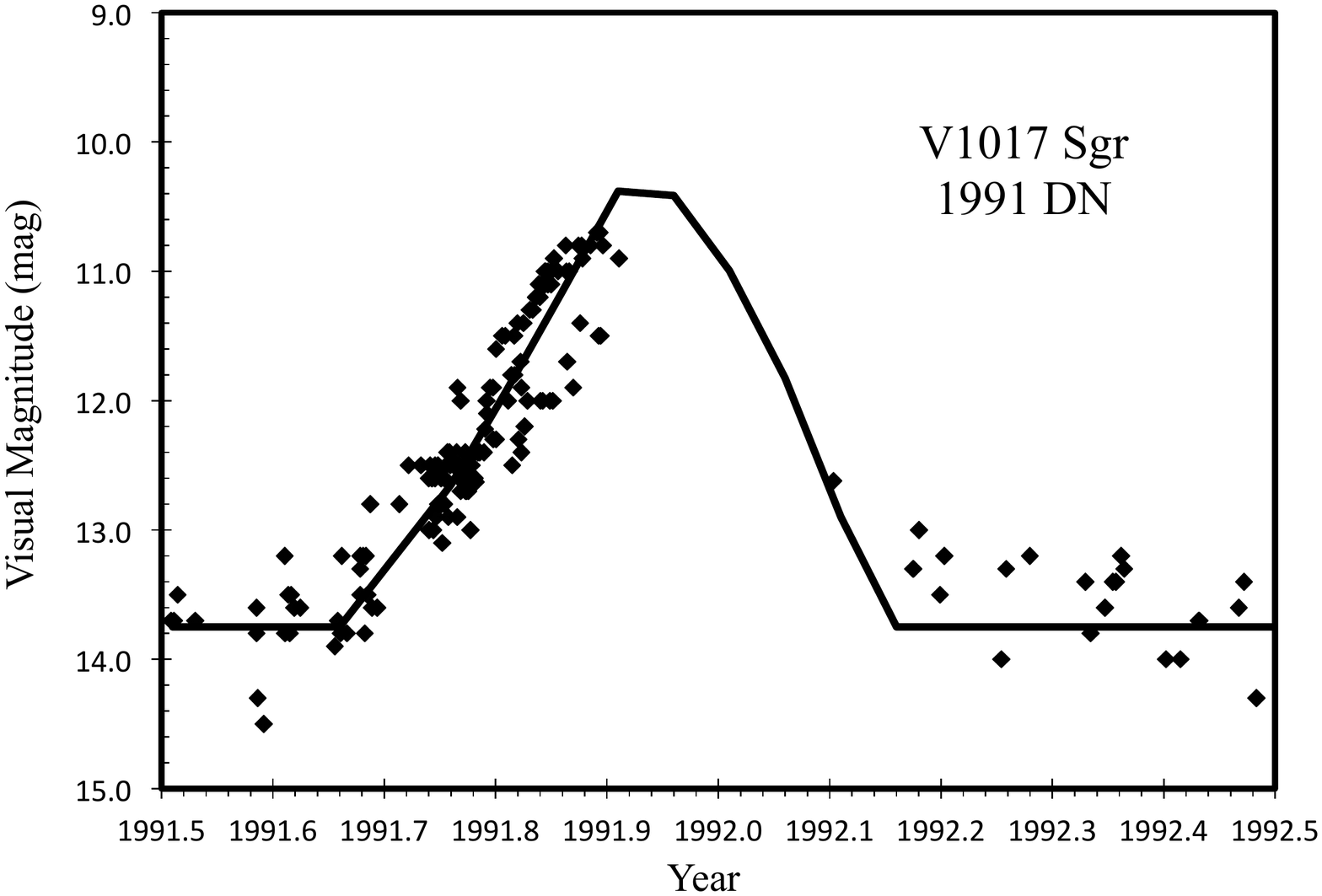}
    \caption{DN light curves for V1017 Sgr.  The 1901 DN light curve is from Harvard archival plates of the B and AM series, corrected for the combined light of CD-29\degr 15053.  The 1973 and 1991 DN light curves are visual magnitudes from the AAVSO.  A template light curve constructed from the binned visual light curves is displayed.  The peak brightness is near 10.4 with an amplitude near 3.3 mag, with a roughly symmetric light curve of total duration 0.5 years.  To within the uncertainties, all three DN events have the same light curve shape.}
    \label{fig:Fig. 3}
\end{figure}

Our light curve has extensive measures of the quiescent brightness levels from 1897 to 2016, and these are good for answering a variety of questions of current interest.  The B and V light curves (from Table 3) are shown in Figure 4, with the points from 1901, 1919, 1973, and 1993 excluded as being not in quiescence.  Furthermore, we display a broken line that represents the trend of the light curve.

Did V1017 Sgr change its quiescent brightness level across the 1919 CN eruption?  Schaefer \& Collazzi (2010) identified eight nova systems for which the post-eruption magnitude level ($m_{post}$ long after the eruption was over) was 2.5 mags brighter than the pre-eruption magnitude ($m_{pre}$).  These are called the `V1500 Cyg novae', and are somehow related to the nova event driving a long-running high accretion rate from the companion star by some mechanism involving the irradiation of the companion by the nova event.  Collazzi et al. (2009) have examined $m_{pre}$ and $m_{post}$ for 30 nova systems, finding that one-in-six are V1500 Cyg stars.  For V1017 Sgr, the HCO A plates gives $m_{pre}$=15.09 for 16 plates from 1897 to 1916, and $m_{post}$=14.60 for 28 plates from 1923 to 1950.  This gives $m_{pre}-m_{post}$= +0.49 mag, which is within the normal range of variation for stars with little irradiation effect.  With a long orbital period, we expect that irradiation effects will be minimal.

Did V1017 Sgr display any rise or dip in the light curve immediately before the eruption?  Collazzi et al. (2009) have found three nova systems that display rises and dips on time scales from a month to a decade, while Schaefer et al. (2013) found a significant rise starting ten days in advance of the 2011 T Pyx eruption.  For the most similar case, T CrB (another CN from a system with a giant companion) has a one magnitude rise starting a decade before its 1946 CN event.  For V1017 Sgr, we have no resolved images from 1916 to 1918, so short duration rises or dips cannot be recognized.  However, the quiescent light curve shows a clear, significant, and steady rise from around B=15.3 in 1907.5 to B=14.7 in 1916.5.  The case for a causal connection between the rise and the 1919 CN event is not strong, because we see long-term fluctuations at the 0.5 mag level (like the drop from 1977 to 1990) at other times.

Did V1017 Sgr show a systematic decline, as predicted by the 'Hibernation' model (Shara et al. 1986), caused by the binary system slightly pulling apart due to the mass-loss from the CN eruption?  We now have four-out-of-five CN for which the orbital period has been measured to decrease across the CN eruption, so this shows that most novae are not pulling apart due to an eruption.  Other ideas involving irradiation still lead to the decline in brightness of a system long after the CN eruption is over (e.g., Patterson et al. 2013).  For V1017 Sgr, from Figure 4, in the B band, we see that the light curve was flat from 1923 to 1977.  There is a significant decline of about 0.5 magnitude from 1977 to 1990, but this is too fast to be any hibernation or irradiation effect.  In any case, the light curve recovered back to its 1977 level by 1993.  From the V light curve, we see that the magnitude dropped by perhaps 0.3 mag from 1977 to 2015.  None of these variations are as predicted or expected from any hibernation or irradiation effects.

How can V1017 Sgr possibly have long-term changes in its quiescent level at the 0.3 to 0.6 mag level?  In ordinary CVs, the optical brightness is dominated by the light from the accretion disk, so it is easy to make the quiescent level change simply by varying the accretion rate.  We have no good understanding why accretion rates can change on any time scale, and such variations are common on all time scales, so it is easy to make this required connection.  But this easy answer cannot apply to V1017 Sgr because the optical light is greatly dominated by the companion star, so it would take very large variations in the accretion rate to explain the modest changes as observed.  Furthermore, our I band magnitudes show the same drop from 1977 to 1990 with the same amplitude as in the B band, so the long term light variations in quiescence are certainly not coming from the accretion disk.  The only possibility is that the brightness of the companion star is changing by 30\% to 50\%.  This is startling, since red giants confined in Roche lobes should not have secular changes in brightness on time scales of a dozen years.  The best idea that we can think of is that the accretion rate changes, so the irradiation of the companion by the accretion disk change, so the surface temperature of the companion changes, thus making the brightness level of the companion vary up and down.

\begin{figure}
	\includegraphics[width=1.1\columnwidth]{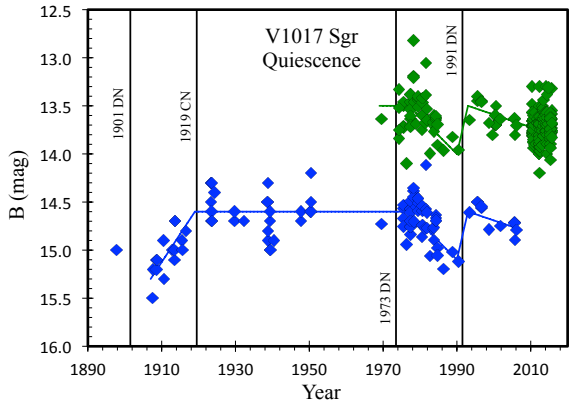}
    \caption{V1017 Sgr in quiescence.  This light curve is from all data 1897-2016, excluding magnitudes from 1901, 1919, 1973, and 1991.  All these magnitudes are in the modern Johnson B system (the blue points) and in the Johnson V system (the green points).  The points to take from this plot are (1) there is a pre-eruption rise by $\sim$0.5 mag over a dozen years, (2) there are long term fluctuations at the one-third of a magnitude level during quiescence, and (3) there is no systematic decline in the quiescent brightness level after the 1919 CN eruption.}
    \label{fig:Fig. 4}
\end{figure}

\section{Post-Eruption Orbital Period}

We used our light curve to search for a photometric periodicity. In part, this is to test the period of 5.714 days suggested by Sekiguchi. Sekiguchi reported this orbital period with just a set of 9 radial velocity measurements over a 37 day interval of observation, and this data set is subject to many aliases. With our much larger data set over a very long time span, we should be able to determine the photometric period with reliability, with high accuracy, and without ambiguities.

For our analysis, we ran a Discrete Fourier Transform (DFT) on the data. The Fourier Transform is optimal for the near-sinewave folded light curves that we find, and our sparse data requires that we use the `discrete' version of the Fourier Transform.  Furthermore, once the best period was identified for each data set, we then performed a chi-square fit to a sine wave.  This fit gives us the best fit epoch, the best fit amplitude, and it gives us exact measures of the error bars by a well-known and robust method.

The best data set for period searches is that of Dvorak in the V-band, because there are a large number of nights over just a few years.  The DFT shows a highly significant period at 2.89317 days.  (There are also daily alias peaks at 0.7416 day and 1.5218 day.  These aliases are rejected from the many other data sets.)  Further, there is no significant power at twice the period.  This period will be confirmed with several other data sets.  So we have our answer: the photometric period of V1017 Sgr is $P_{phot}$=2.89317 days.

However, there is a possibility that the period that we are seeing photometrically is actually half of the orbital period because of ellipsoidal variations caused by the large companion star in our system. Since the companion star is dominating the optical light in quiescence it is reasonable to expect that ellipsoidal variations will dominate in the light curve. If so, then $P_{orbit}$=2$\times P_{phot}$. 

With this period-doubling, we have an orbital period of 5.78634 days.  This is right in the range suggested by Sekiguchi.  Sekiguchi's radial velocity data are completely inconsistent with any period from 1.5 to 3.5 days.  Thus, the true orbital period is close to the original Sekiguchi period of 5.714 days, meaning that Sekiguchi was right.

The Dvorak data folded on the DFT photometric period shows an ordinary sine wave shape.  We have fitted a simple sine wave.  The epoch is taken to be the epoch of minimum brightness, so that this can be compared to Sekiguchi's epoch (which corresponds to the time of the superior conjunction of the companion star).  The epoch is also selected for a time close to the average time of the data set, so that the period and epoch error bars have minimal correlation.  With this, we give the full best fit parameters in Table 4.

\begin{table*}
	\begin{minipage}{200mm}
	\centering
	\caption{Chi-Square Fits to Sine Waves.}
	\label{tab:table4}
	\begin{tabular}{llllllllll} 
		\hline
		Fit description & Band & Years & \# & Epoch (JD) & Period (day) & $\dot{P}$ (day/day) & Amp. (mag) & $\sigma$ (mag) & $\chi^2$ \\
		\hline
Dvorak only	&	V	&	2010-2015	&	139	&	2456229.292	$\pm$	0.067	&	2.893199	$\pm$	0.000304	&	0		(fixed)	&	0.230	$\pm$	0.040	&	0.14	&	130.13	\\
All B (post-eruption)	&	B	&	1923-2015	&	82	&	2439897.496	$\pm$	0.099	&	2.893139	$\pm$	0.000031	&	0		(fixed)	&	0.199	$\pm$	0.046	&	0.14	&	97.49	\\
All V (post-eruption)	&	V	&	1969-2015	&	215	&	2453116.187	$\pm$	0.058	&	2.893212	$\pm$	0.000033	&	0		(fixed)	&	0.214	$\pm$	0.027	&	0.14	&	208.10	\\
All R (post-eruption)	&	R	&	1973-2006	&	63	&	2449227.692	$\pm$	0.137	&	2.893025	$\pm$	0.000083	&	0		(fixed)	&	0.163	$\pm$	0.051	&	0.14	&	56.47	\\
All (post-eruption)	&	BVR	&	1923-2015	&	360	&	2449375.375	$\pm$	0.050	&	2.893145	$\pm$	0.000016	&	0		(fixed)	&	0.192	$\pm$	0.021	&	0.14	&	369.71	\\
All (post-eruption) with $\dot{P}$	&	BVR	&	1923-2015	&	360	&	2449375.319	$\pm$	0.049	&	2.893167	$\pm$	0.000015	&	(1.19$\pm$0.54)$\times$10$^{-8}$			&	0.194	$\pm$	0.020	&	0.14	&	368.16	\\
Pre-eruption	&	B	&	1907-1916	&	15	&	2419332.771	$\pm$	0..068	&	2.893849	$\pm$	0.000165	&	0		(fixed)	&	0.412	$\pm$	0.073	&	0.09	&	12.14	\\
Pre-eruption, $\dot{P}$=$\dot{P}_{post}$	&	B	&	1907-1916	&	15	&	2422018.064		(fixed)	&	2.893694	$\pm$	0.000174	&	1.19$\times$10$^{-8}$		(fixed)	&	0.412		(fixed)	&	0.09	&	12.76	\\
Final joint fit post-eruption	&	BVR	&	1923-2015	&	360	&	2422018.203	$\pm$	0.155	&	2.893019	$\pm$	0.000039	&	(1.6$\pm$0.6)$\times$10$^{-8}$			&	0.194	$\pm$	0.020	&	0.14	&	366.29	\\
Final joint fit pre-eruption	&	B	&	1907-1916	&	15	&	2422018.203	$\pm$	0.155	&	2.893808	$\pm$	0.000136	&	(1.6$\pm$0.6)$\times$10$^{-8}$			&	0.412	$\pm$	0.077	&	0.09	&	12.28	\\
		\hline
	\end{tabular}
	\end{minipage}
\end{table*}

The RMS variation of the magnitudes ($\sigma$) around this best fit is greatly larger than the photometric uncertainty.  This observed scatter is from the usual flickering suffered by all CVs.  So the 1-sigma uncertainty for use in a chi-square fit is the addition in quadrature of the small measurement error ($\sigma_{meas}$) plus the star's intrinsic variation from flickering ($\sigma_{flicker}$), with $\sigma^2=\sigma_{meas}^2+\sigma_{flicker}^2$.  This means that the photometric uncertainty for CCD and photoelectric measures is always negligibly small to the usual flickering noise in the star.  We can only get an empirical measure of the intrinsic variance of the star, and this is simply chosen so that the reduced chi-square is near unity.  This procedure means that the reduced chi-square cannot be used as a criterion for the quality of the fit, but {\it changes} in the chi-square are valid measures for estimating error bars and relative confidence levels.

For the Dvorak data, the fit has a minimum chi-square of 130.13 with 135 degrees of freedom.  The one-sigma uncertainties are with the parameters within 1.0 of this minimum.  If we set the sine wave amplitude to zero, then the chi-square is 178.6.  That is, the chi-square difference is 48.5, which is to say that the existence of the periodicity is 7.0-sigma.

Some of our other various data sets have the problem that they cover a many-year time span, over which secular variations move the average level up and down by amounts comparable to the amplitude.  This makes for large noise in the DFT due to folding trying to line up the magnitudes taken in the faint state, all masking the real period.  To solve this, we have subtracted out the long term trend, as shown in Figure 4.  The detrended light curve hovers around zero, all with a near-zero mean.  This step is the only way to spot the photometric periodicity in the very long data sets that we have.

With this, we can spot several outlier magnitudes, with these being isolated and $>$3-sigma away from any folded light curve.  These are on 2442904 (from Walker), 2443608 (from the CTIO 0.4-m), and 2444826 (from the CTIO 4-m).  These are just the usual outliers, and it is not clear whether these are some observational error (e.g., thin clouds) or whether V1017 Sgr was simply variable above the periodic photometric oscillation (e.g., with an unusually large flicker).  We have rejected these outliers, as otherwise the DFT would have substantial noise as they beat against each other.

With the detrending and the outlier rejection, the same photometric periodicity is significant in the data sets from Landolt for each filter individually, from HCO, and from SMARTS.  This gives us high confidence that we have found the real photometric period.

Next we performed fits for all the post-eruption B, V, and R data individually.  These fits are reported in Table 4.  We see that the fitted amplitude does not change significantly with band.  This is consistent with normal ellipsoidal variations, where the amplitude just comes from the shape of the star in all colors.  Furthermore, we see that the reduced chi-square is near unity for all of the colors having the same $\sigma$, and this implies that the flickering amplitude is roughly a constant with color.  Normal flickering light in CVs is very blue in color, so the intrinsic variations of the star are apparently from some other mechanism.

The fits for the three colors are fairly consistent in period, with an average near 2.89316 days.  The amplitudes are consistent with having no color dependence, with an average value of near 0.20 mag.  There is no significant evidence that the RMS scatter around the best fit light curves (0.14 mag) has any color dependence.  With no color dependence in the detrended light curves, the best and final values will come from taking all detrended magnitudes after the 1919 CN eruption.

For all 360 B, V, and R magnitudes from 1923-2015, the DFT returns a single dominant period of 2.893144 days.  Our chi-square fit (see Table 4) returns the same period.  The period is accurate to 5 parts-per-million.

We can fold all the magnitudes to see the shape of the light curve.  To see the folded light curve in quiescence, we have not included any magnitude from 1901, 1919, 1973, or 1991.  With the star's quiescent level varying around on decadal time scales, the scatter in a straight fold would be large, so we have used the detrended light curves instead.  We do not have enough data from any one band to make a clean light curve, so we have combined the detrended magnitudes from B, V, and R bands.  Like for many CVs, the star's intrinsic flickering makes for scatter that is comparable to the ellipsoidal effects, with this serving to hide the underlying variations.  The solution is to phase average the magnitudes, and we will do this in bins of size 0.05 in phase.  With this, the folded and phase-binned light curves for V1017 Sgr in B and V are presented in Figure 5 for a period of 2.893145 days.

\begin{figure}
	\includegraphics[width=1.1\columnwidth]{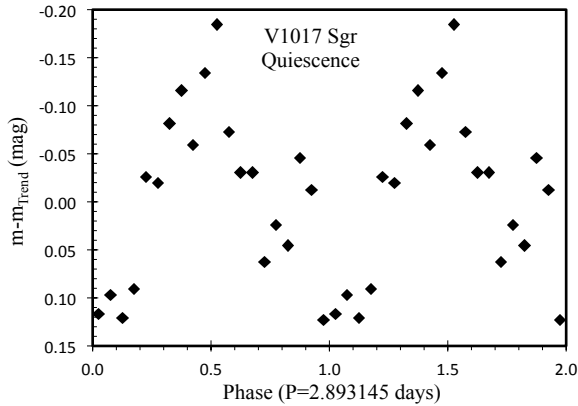}
    \caption{V1017 Sgr folded light curve for $P$=2.893145 days.  This light curve is for all post-eruption detrended magnitudes in B, V, and R, all averaged within 0.05 phase bins.  We see an apparent sine wave with a period equal to half the orbital period, pointing to ellipsoidal variations.  Such is expected when the optical light curve is dominated by the companion star.}
    \label{fig:Fig. 5}
\end{figure}

Given Sekiguchi's radial velocity curve and the likelihood that the light curve will be dominated by the companion's ellipsoidal variations, the orbital period is twice the photometric period, or $P$=5.786290$\pm$0.000032.  

CVs have mass transfer and other effects that make for a slow and steady period change ($\dot{P}$ measured in units of days per day) while in quiescence between nova eruptions.  We have found that such a steady period change can be quite large for the recurrent nova T CrB, which also has a red giant companion star and CN eruptions.  So we have run further fits with all the 360 de-trended post-eruption magnitudes in B, V, and R (see Table 4).  We do find a marginally significant steady period change.  While the existence of the $\dot{P}$ is not required, it is expected and it is the best representation of the data.  With this, the times of the photometric minima are JD $2449375.319 + E\times 2.893167 + (\frac{1}{2} E^2 1.19\times10^{-8}$), for E being an integer.  Around the start of March 1919, the time of minimum is JD 2422018.064$\pm$0.150.

\section{Pre-Eruption Orbital Period}

    Our group has been vigorously pursuing the measurement of pre-eruption orbital periods ($P_{pre}$) of novae, so that, when combined with accurate post-eruption orbital periods ($P_{post}$), we measure the {\it change} of the period across the eruption ($\Delta P$=$P_{post}-P_{pre}$).  This measure provides an accurate model-free measure of the mass ejected by the nova event (an otherwise very-poorly known quantity), with implications for nova physics and for models of CV evolution (this alone refutes the Hibernation model).  Now, for V1017 Sgr, we have just derived a parts-per-million post-eruption orbital period, and we have HCO data that shows the orbital variations from 1907 to 1916.  (The magnitude from 1897 cannot be used because we do not know the trend for detrending.)  So we have tried to pull out the pre-eruption orbital period.
    
Given the substantial photometric modulation on half the orbital period visible from 1923-2015, we can strongly expect that V1017 Sgr should show substantial photometric modulation on half the orbital period from 1907-1916.  For the 15 B-band magnitudes from 1907-1916, the DFT shows many nearly equal peaks, as appropriate for the small number of points.  These many peaks are simply caused by the high points and low points randomly beating each other to produce spurious aliases over a very wide range of periods.  These aliases are certainly not physical, and we can recognize this by them requiring an impossibly large period change across the 1919 CN eruption.  Based on theory (see Schaefer 2011) plus our many well-measured $\Delta P$ values (now for 5 CN eruptions and for 4 recurrent nova eruptions), we know that $\Delta P/P$ must be smaller than a 0.1\% change.  For V1017 Sgr, this means that any apparent periodicity outside the range 2.890--2.896 days must be a spurious alias. We are strongly expecting a significant DFT peak inside the period range 2.890--2.896 days that is the real half-period of V1017 Sgr.

The DFT of the 15 HCO pre-eruption de-trended B magnitudes has a high peak at a period of 2.893877 days.  The chi-square fit to a sine wave for this Fourier peak is given in Table 4.  We see that the amplitude is larger than the post-eruption amplitude, and the RMS scatter around this best fit is somewhat smaller.  We show the folded light curve in Figure 6, and it looks good.  We can test the significance of the sine wave periodicity by comparing the chi-square values for the best fit case (12.14) versus the zero-amplitude case (43.96).  The difference in chi-square is 31.82, and this corresponds to a 5.6-sigma confidence level in the significance of the periodic modulation.  So even with only 15 magnitudes with tenth-magnitude variations, we have a highly significant photometric modulation at the half-orbital period.

\begin{figure}
	\includegraphics[width=1.1\columnwidth]{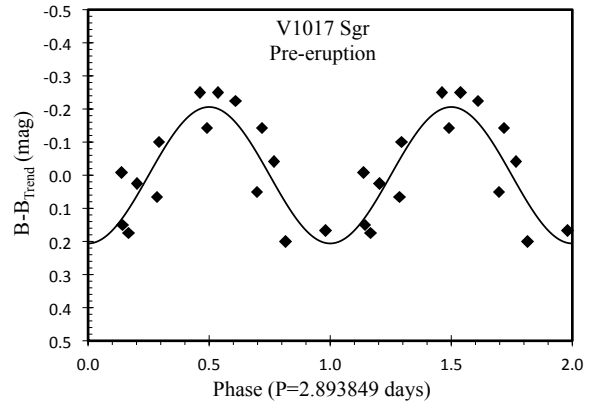}
    \caption{V1017 Sgr folded light curve for pre-eruption data.  This light curve is for the 15 pre-eruption HCO magnitudes in B.  We see what looks like a good sine wave, pointing to this being a good photometric period.  From the chi-square analysis, the existence of this photometric periodicity is at the 5.6-sigma confidence level.}
    \label{fig:Fig. 6}
\end{figure}

The change in the photometric period is -0.000704$\pm$0.000170 days, while the change in the orbital period is -0.001408$\pm$0.000340 days.  That is, the orbital period got {\it shorter} across the 1919 CN eruption by $\Delta P/P$=-243$\pm$57 parts-per-million.  This large of a change in the orbital period is startling for theory, so we have to carefully test whether the orbital period change across the eruption is significant, and we have to run other tests as we can.

One test is to see whether the pre-eruption ephemeris extrapolates to a yield an epoch around the 1919 eruption that is consistent with that derived from the post-eruption ephemeris.  That is, the period cannot have any jumps across the eruption, which is to say that the $O-C$ curve must be continuous.  From the post-eruption fit, we derived a minimum time of JD 2422018.064$\pm$0.150.  From the pre-eruption fit, we derive an epoch of JD 2422018.263$\pm$0.168.  The time difference is 0.199$\pm$0.225.  That is, the the pre-eruption ephemeris does well match the post-eruption ephemeris to within 1-sigma.  If the pre-eruption period is spurious, then we would expect a random 84\% chance that there would be a mismatch.  Thus, this test provides a small amount of confidence in the period.

A second test is to examine the fit and folded light curve for the 15 pre-eruption magnitudes as based on the post-eruption ephemeris.  The folded light curve is shown in Figure 7.  The post-eruption ephemeris is obviously horrible.  The poor folded light curve is not just a matter of shifting the epoch, as there is a large RMS scatter at all phases, and this is the symptom of a bad period.  This demonstrates that there must be some significant period change across the 1919 CN eruption.

\begin{figure}
	\includegraphics[width=1.1\columnwidth]{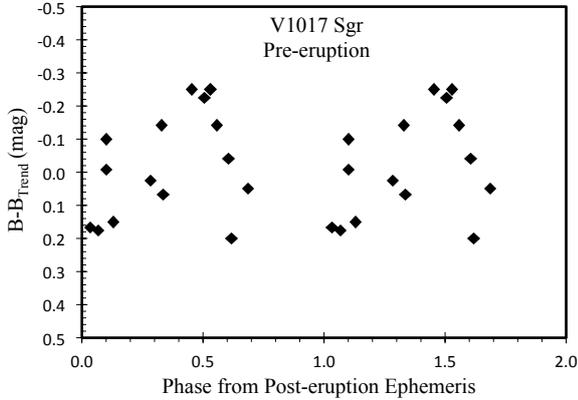}
    \caption{V1017 Sgr pre-eruption data folded on the post-eruption period.  This light curve is the same as for Figure 6, except that the {\it post}-eruption period is used for the fold.  Now, we see that the nice sine wave goes away.  This demonstrates that the pre-eruption period is distinctly different from the post-eruption orbital period.}
    \label{fig:Fig. 7}
\end{figure}

A third test is to see whether the 1897 measure fits into this best fit.  The magnitude in 1897 was not de-trended because it is isolated in time at a decade before any other magnitude, so we have no confident and independent way to know how the trend goes in 1897.  Still, looking at the B magnitudes on the far left of Figure 4, any plausible trend line would fall below, likely substantially below the 1897 measure.  This would give a weak test that the the phase of the 1897 point should be in the bright half of the curve (i.e., phase between 0.25 and 0.75), and should likely be near the phase 0.5 around the peak brightness.  The phase of the 1897 point is 0.62, which agrees with the expectation that the period is real.  At worst, this test is only a 50\% test, yet with the period having passed this weak test, we still do get some small amount of extra confidence.

The linear fit for the pre-eruption data was presented as a simple case to define the problem.  We should perform fits that make use of two more physical facts that we have established from the post-eruption data.  First, we know from the post-eruption fit that the pre-eruption ephemeris must pass through JD 2422018.064$\pm$0.150.  Second, we should be using the $\dot{P}$ in the pre-eruption fit, and this $\dot{P}$ value should be the same as that for the post-eruption fits.  This new fit would set the epoch to JD 2422018.064, $\dot{P}$ to 1.19$\times$10$^{-8}$ days/day, and keep the amplitude at 0.412 mag.  This fit is summarized in Table 4.  The folded light curve is virtually identical to that in Figure 6 except with a 0.04 phase shift to the right.  The pre-eruption period has changed substantially from the post-eruption period.  With the $\dot{P}$, we have to be careful to compare the periods immediately before and after the eruption.  With this, $P_{pre}$=2.893685$\pm$0.000060, $P_{post}$=2.893054$\pm$0.000015, and $\Delta P$=0.000631$\pm$0.000062 days.  This corresponds to a 10.2-sigma confidence level for the period having suffered a change across the CN eruption.

So far, our fits have treated the time intervals before and after 1919 (mostly) independently.  But the physical constraints of the situation forces the lack of any jumps in the epochs of zero-phase, and this means that the pre- and post-eruption intervals should be connected with a joint fit.  In an $O-C$ curve and in the ephemeris of superior conjunction, this can be represented as a broken parabola.  That is, there will be one epoch around the start of March 1919, the ephemeris after 1919 will be some parabola, and the ephemeris before 1919 will be a different parabola.  The $\dot{P}$ before the eruption is poorly constrained by the data, so we will simply presume the obvious default case that the steady period change remains constant over all times in quiescence.  With this, we then have four fit parameters for the period; the epoch of a superior conjunction around the start of March 1919, the post-eruption orbital period, the pre-eruption orbital period, and $\dot{P}$.

Our final fit will be one over-all fit, from 1907 to 2015, with de-trended magnitudes for B, V, and R.  As one joint fit, we ensure that the connection in 1919 is made exactly.  We have 375 magnitudes.  The final over-all joint fits are presented in the last two lines of Table 4.  These best fit parameters are for the photometric periodicity, with the epoch being the time of the star at minimum brightness.  The 1-sigma error bars are calculated as the maximal deviation from the reported best parameter, for any combination of the other parameters, for which the resultant chi-square is 1.00 larger than the minimum chi-square.  

The orbital periodicity is twice the photometric periodicity, while the $\dot{P}$ is four times larger.  To be explicit, here are the pair of ephemeris equations ,
\begin{equation}
JD= 2422018.203 + 5.786038 E + 0.5 E^2 6.4\times10^{-8}, E>0,
\end{equation}
\begin{equation}
JD= 2422018.203 + 5.787616 E + 0.5 E^2 6.4\times10^{-8}, E\leq0, 
\end{equation}
where $E$ is an integer.  The uncertainties on these coefficients can be taken from Table 4, although the uncertainty for the periods are doubled and $\dot{P}$ are quadrupled.  The epoch in these equations corresponds to the time of the deepest of the two minimum in the folded light curve (cf. Figure 8).  

The orbital phase for any time will be the fraction of the interval between two adjacent epochs of zero phase.  These orbital phases are listed for all observations in Table 3.  These orbital phases have also been used to create a folded light curve, as shown in Figure 8.  We see photometric maxima near phase 0.25 and 0.75.  Photometric minima are near phase 0.00 and 0.50, with the zero-phase minimum being significantly deeper.  This folded light curve is characteristic of ellipsoidal modulation with some sort of an eclipse at near zero phase.

\begin{figure}
	\includegraphics[width=1.1\columnwidth]{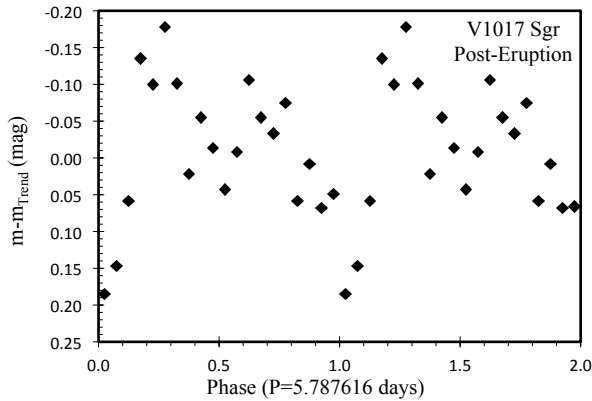}
    \caption{V1017 Sgr folded light curve for $P$=5.787616 days.  This light curve is for all post-eruption detrended magnitudes in B, V, and R, all averaged within 0.05 phase bins.  The 1-sigma error bars of each point are roughly 0.03 mag.  We see an ellipsoidal modulation on the orbital period, with the minimum at 0.0 phase being deeper than at 0.5 phase.  Apparently, the accretion disk is partially eclipsing the companion star.}
    \label{fig:Fig. 8}
\end{figure}

The critical point is that is that the photometric modulation has $P_{pre}$=2.893808$\pm$0.000136 days (for just before the 1919 eruption), $P_{post}$=2.893019$\pm$0.000039 days (for just after the 1919 eruption), and so $\Delta P$= -0.000789$\pm$0.000142.  With the 1-sigma range in parameter space being moderately correlated, the real 1-sigma error bar for $\Delta P$ is moderately larger, so $\Delta P$= -0.000789$\pm$0.000176.  The smallest chi-square for parameters that can return a zero period change is 32.89 larger than the minimum, which points to the period change as being 5.7-sigma.  In all cases, the existence of the period change is highly significant.  The period has changed across the 1919 CN eruption with $\Delta P/P$ equal to -273$\pm$61 parts-per-million.  The orbital period {\it decreased} across the 1919 CN eruption.

This result is startling for nova theorists.  The reason is that the change in period should be comparable to  $P\times(M_{ejecta}/M_{WD})$ (Prialnik and Shara 1986, Schaefer 2011), and this should be greatly smaller than the observed $\Delta P$.  Furthermore, for V1017 Sgr, the ejection of mass can only make for a period {\it increase}.  So we have a problem, at least to the extent that modern nova theory is missing some large evolutionary effect, having no glimmer of understanding for large period decreases.

This huge period {\it decrease} is {\it not} startling to our group here at Louisiana State University, because we have now measured $\Delta P/P$ across 5 other {\it classical} nova eruptions, and three of those five also had huge period decreases.  In particular, RR Pic, HR Del, and QZ Aur had $\Delta P/P$ values between -288 and -451 parts-per-million shortenings of the period across their CN eruption, and these changes are all measured with high confidence and high accuracy.  (DQ Her also suffered a highly significant period decrease, but only at the 4.8 parts-per-million level.)  The CN BT Mon has an observed period increase across its 1939 eruption (Schaefer and Patterson 1983), and this was part of the original motivation for the 'Hibernation' model and this also shows that not all novae have their period decreasing. So, our new result for V1017 Sgr is merely the fifth case (out of six measured) with a period {\it decrease}, and is the fourth case (out of six measured) with a period {\it decrease} by more than 200 parts-per-million.

We only have 15 useable pre-eruption magnitudes, each with a photometric accuracy of around 0.09 mag, so some willful researchers might think to deny or ignore our result.  Such would be the easiest way to solve the theory problem.  Still, despite the low numbers of pre-eruption plates, our period change is of high confidence, and has passed a variety of additional tests:  (1) Our chi-square fits show that the standard significance of the existence of a sinusoidal modulation is at the 5.6-sigma confidence level in the pre-eruption data.  (2) The folded light curve for the pre-eruption data looks good in Figure 6.  (3) The freely-fit pre-eruption ephemeris crosses the post-eruption ephemeris close to the time of the 1919 CN event.  (4) The pre-eruption light curve folded with any small period change looks horrible, as shown in Figure 7.  (5) The 1897 magnitude measure easily fits in to the bright part of the folded light curve.  (6) The lowest chi-square for any case with a near-zero period change is 32.89 larger than the minimum chi-square, so the period change by this measure is significant at the 5.7-sigma level.  (7) A huge period decrease is actually the case for three-out-of-five CN events, so V1017 Sgr making it four-out-of-six is reasonable.  In all, we have a very strong case that V1017 Sgr had a -273$\pm$61 parts-per-million period decrease in 1919.

For the work reported in this paper, we originally started out just trying to test Sekiguchi's reported period, and to get a good photometric light curve.  Indeed, we have confirmed Sekiguchi's period, although we can offer greatly higher accuracy and reliability.  With our collected data, we found that we can measure the post-eruption orbital period to parts-per-million accuracy.  With this, we then realized that we actually have enough pre-eruption data that we can measure the photometric modulation and determine the orbital period to useful accuracy.  Startlingly (for theorists), we find that the orbital period has decreased across the 1919 CN eruption by -273$\pm$63 parts-per-million.  This is a result where the change in period is significant at the 5.7-sigma confidence level and has passed various tests, so we consider it strongly proven.  This is just our fourth example of a huge period decrease across CN eruptions, so it is only surprising for nova theorists.

\section{Stellar Masses and Radial Velocities}

For various analysis questions, we need to know the mass of the white dwarf ($M_{WD}$) and the mass of the companion star ($M_{comp}$), and these are tied in to the radial velocity curve of Sekiguchi (1992).  We have a variety of constraints and results:

With our new ephemeris, we have phased up Sekiguchi's nine radial velocity measures and fit them to a sine wave.  We find a systemic velocity of 15 km/s and an amplitude of 90 km/s.  The epoch when the GIII companion star (for which the lines are measured) is at superior conjunction (i.e., the accretion disk is directly in front of the companion) is when the star's velocity is zero (relative to the systemic velocity) and running negative (i.e., the companion starts to come towards the Earth).  From our fit, the phase of superior conjunction is 0.03.  The error bars on these fit values are unclear because the fitted sine wave has a very large reduced chi-square, so the real uncertainties must be greatly larger then tabulated by Sekiguchi.

With our period, we calculate that the system has a mass function of 0.44 M$_{\odot}$.  This is a lower limit on $M_{WD}$.  The mass function depends on $M_{WD}$, the mass ratio $q$ ($M_{comp}$/$M_{WD}$), and the orbital inclination ($i$).  $M_{WD}$ must be larger than 0.6 M$_{\odot}$ for a field white dwarf, and should be less than something like 1.2 M$_{\odot}$ because no recurrent nova events are seen in 119 years despite a high apparent accretion rate.  Our mass function is a large fraction of any $M_{WD}$  value, so that means that the inclination must be fairly edge-on and the mass ratio cannot be too large.  Extreme limits on the various parameters are $M_{WD}>$0.45 M$_{\odot}$, $i>$46\degr, and $q<$0.66.  A low $M_{comp}$ is incompatible with the companion being the size of a yellow giant star, so we expect that there must at least be some sort of a normal core, so $M_{comp}>$0.6 M$_{\odot}$ or so.  There will then be a limit on the mass ratio, where $q>$0.5 or so.  With this, we are forced into the white dwarf being near the maximal mass and the inclination being near edge on (i.e., eclipsing).  With our lower limit on the companion mass, we have that the the white dwarf must be $>$1.07 M$_{\odot}$ and that the inclination must be $>$70\degr.  For the extreme case with $M_{WD}=$1.2 M$_{\odot}$ and $i$=90\degr, we have $M_{comp}$=0.79 M$_{\odot}$, and $q$=0.66.  For a middle-of-the-range case with $M_{WD}=$1.1 M$_{\odot}$ and $i$=80\degr, we have $M_{comp}$=0.60 M$_{\odot}$, and $q$=0.55.  

We can now derive the system geometry, and we'll adopt the middle-of-the-range case.  We find that the separation between the centers of the companion star and the white dwarf is 16.1 R$_{\odot}$, and that the radius of the companion star's Roche lobe is 5.3 R$_{\odot}$.  This means that the companion star will eclipse the central white dwarf for inclination $>$72\degr, while it will eclipse parts of the accretion disk to much lower inclinations.  This means that in all cases we should see some sort of an eclipse.  And this is consistent with the light curve seen in Figure 8.

We have found that the phase of the superior conjunction of the companion star is close to the phase of the deepest of the two minima in the folded light curve (see Figure 8).  This means that the system is dimmed more by the disk covering up the companion star than by the companion star covering up the same area of the disk.  For detached eclipsing binaries, the rule is that the deeper minima should be associated with the hotter star being eclipsed.  For V1017 Sgr, the companion star has a surface temperature of 5200 K, while the outer parts of the disk must have a typical temperature of $\sim$10,000 K (so that DN events can occur) and the inner parts are even hotter.  So we do not understand the depths of the minima in the folded light curve.

The companion star (5200 K and 5.3R$_{\odot}$) is certainly an evolved star above the main sequence.  We are not clear whether it should be more correctly called a sub-giant star or a giant star.  In all cases with no binary interaction, the progenitor would have to come from a star with a main sequence mass of roughly 2 M$_{\odot}$ to get to that position on the HR diagram.  Further, the progenitor must have spent most of its main sequence lifetime with a mass $>$1 M$_{\odot}$ so as to start evolving off the main sequence in the age of the Milky Way.  But there must have been binary evolution when the now-white-dwarf star became a giant with a size larger than the current binary system.  This interaction could have substantially tightened the orbit, reducing the orbital period down to 5.78 days.  This interaction inevitably must have involved some substantial mass transfer onto the now-companion star.  During the CV phase, mass will transfer from the companion to the white dwarf, with a lot of this material then being blown off as ejecta from the CN eruptions.  As above, the companion star appears to be somewhere between 0.6 M$_{\odot}$ and 0.79 M$_{\odot}$.  So apparently, the companion has net lost something like 0.2 M$_{\odot}$ or more during its complex binary evolution.

\section{Distance to V1017 Sgr}

We can now estimate and limit the distance, $D$, to V1017 Sgr:

With our SED and $P$, we can derive a blackbody distance to the giant companion.  Previously, we have done this with two independent methods for five recurrent novae (Schaefer 2010).  Both methods key off the size of the companion star, as derived from the orbital period.  The radius of the companion must equal to the radius of its Roche lobe, and this comes from Kepler's Law and the usual Roche lobe geometry.  For this, there is a weak dependence on the star masses.  We adopt 1.1 M$_{\odot}$ for the white dwarf and 0.6 M$_{\odot}$ for the companion star.  Then, with a 5.786 day period, the companion star has a radius of 5.3 R$_{\odot}$.

Our first method for the blackbody distance works through the absolute bolometric magnitude.  With the usual Boltzmann equation, as scaled from our Sun, the absolute bolometric magnitude of the 5200 K companion with radius 5.3 R$_{\odot}$ is +1.6 mag.  For $E(B-V)$=0.39 mag, the V-band extinction is 1.2 mag.  With the observed V magnitude equal to 13.5 mag, the extinction-corrected V magnitude is 12.3.  For a G5 $III$ star, the bolometric correction is -0.3 mag, so the companion star has a bolometric magnitude of 12.0 mag.  The distance modulus is then 12.0-1.6 = 10.4 mag, which corresponds to 1210 parsecs.  This distance is likely to be slightly on the low side, because the accretion disk must account for some 10\% or so of the V-band flux (so as to account for the flickering).  The uncertainty is dominated by the variation in the V magnitude and the error in determining the companion's surface temperature from the SED taken at different times, for which we estimate an error bar of $\pm$200 pc.  So, our first method gives a $D$=1210$\pm$200 pc.

Our second distance estimate uses a different route to get the blackbody distance of the companion star, this time through the flux at the peak of the SED.  The SED (see Figure 1) peaks at around 1.0 $\mu$m (3$\times$10$^{14}$ Hz) with extinction corrected $f_{\nu,0}$ of near 70 mJy.  The flux at the surface of the star at the peak is determined only by the blackbody temperature from the peak position, and this will suffer geometric dilution as the distance gets further from the star's surface.  This flux will diminish to that observed at a distance of 1270 pc.  By varying the inputs over their plausible ranges, the error bar on the distance is roughly 300 pc.  So our second distance estimate returns 1270$\pm$300 pc.  Our first two distance estimates are really just one, with these two different paths and different inputs all being equivalent, yet it still provides some cross-checking to demonstrate reliability and uncertainties.

Our third distance estimate is simply from the galactic position of V1017 Sgr, with galactic latitude of -9.1\degr and galactic longitude of 4.5\degr.  With coordinates like this, V1017 Sgr looks to be in the inner bulge of our Milky Way, with a distance of around 6000-10000 parsecs.  We all understand that V1017 Sgr could well be outside this range, but $D\sim$8000 pc is certainly the first guess.

Our fourth distance estimate is from the observed systemic radial velocity of +15 km/s (Sekiguchi 1992).  Unfortunately, by being close to the galactic center, this radial velocity is much as expected for stars at all distances along the line of sight.  So this fourth distance estimate method returns no useful information.

Our fifth distance estimate is from the observed extinction with $E(B-V)$=0.39 mag (Webbink et al. 1987).  The total extinction through the whole Milky Way galaxy along that line of sight has $E(B-V)$=0.23 mag (Schlafly \& Finkbeiner 2011).  This means that V1017 Sgr is out past most of the galaxy's dust.  For a galactic latitude of -9.1\degr, and for most of the dust being within 200 pc of the galactic plane, this sets a lower limit of $D>$1200 pc.  This limit has substantial uncertainties, hard to quantify, due to the clumpiness of clouds along the line of sight.

Our sixth distance comes from the peak magnitude for the 1919 eruption.  The distance modulus to the nova can be taken as $B_{peak,0}$-$M_{B,peak}$, where $B_{peak,0}$ is the extinction corrected magnitude at peak brightness and $M_{B,peak}$ is the absolute magnitude at peak brightness in the B-band.  We have an approximate extinction to V1017 Sgr as $A_B=4.1\times E(B-V)$=1.6 mag.  Let us start with an extreme upper limit on the distance, with $M_{B,peak}>-10$ mag (Downes \& Duerbeck 2000; and Kalsiwili et al. 2011) and the faintest possible peak at B=6.4 mag.  With this, the distance is $<$9000 pc.  But V1017 Sgr is certainly far from this limit, partly because only rare nova peak at such a high luminosity and partly because the linear part of the observed light curve shows that the time of peak observed brightness (with $B=6.4$ mag) is far from the peak.  So this distance limit is really $D$$\ll$9000 pc.  For an extreme lower limit on $D$, we can push $M_{B,peak}$ to -6 mag and we can extend the eruption light curve (see Figure 2) early enough in time so that it would be missed due to the Sun only to peak around 2 mag or so.  With this, $D$$\gg$200 pc.  These extreme limits are not useful.  But we can get a reasonable best estimate.  Most nova peak around absolute magnitude -8$\pm$1, while the best matching from S-class light curves (see Strope at al 2010) to Figure 2 suggests that the real peak was 4.5$\pm$1.0 mag.  This gives an extinction-corrected distance modulus of 10.9$\pm$1.4 mag, or a distance of 1500 pc with a range of 800-2900 pc.

In all, only the blackbody distances offer any accuracy or reliability, with the other estimates all being consistent.  Hence, we are concluding that the distance to V1017 Sgr is 1240$\pm$200 pc.

\section{Evolution of V1017 Sgr}

The critical surprise is that we have strong evidence that the orbital period {\it decreased} by -273$\pm$61 parts-per-million across the 1919 CN eruption.  This has many problems and implications.

The key problem is to explain how V1017 Sgr can possibly have such a large period decrease.  Three effects are known that can change the period across a nova eruption (Schaefer 2011, see Appendix B):  (1) The first effect is simply the loss of mass by the white dwarf due to the mass in the ejected shell.  This effect can only {\it increase} the period.  So this cannot explain the observed period decrease.  Further, this effect is much smaller in size than observed.  For example, for a large shell with an ejected mass of $10^{-6}$ M$_{\odot}$, the fractional period change ($\Delta P/P$) will be close to 2 parts-per-million.  (2) The second effect will be from the loss of angular momentum by the companion star as it moved in the `common envelope' of the ejected shell by means of dynamical friction (Livio 1991).  This will necessarily decrease the orbital period, which is at least in the right direction.  However, this will necessarily be a small fraction of the first effect in size.  So the sum of the first two effect must always be to {\it increase} the orbital period (Schaefer 2011, Eqn. 19).  (3) The third known effect is material in the ejected shell being forced to corotate with the white dwarf due to being entrained in its magnetic field for the initial part of its ejection, with this carrying away rotational angular momentum that then gets transferred to cause a loss of the orbital angular momentum (Martin et al. 2011).  This effect has a strong dependence on the mass ratio, where the total period change can be negative only for $q<$0.4 or so, while the required effect (for our observed $\Delta P/P$ and for even $10^{-6}$ M$_{\odot}$ ejected) requires a mass ratio of $q$=0.002 even for high magnetic fields. For V1017 Sgr, we always have $q>$0.5, so this third effect can never make for such a large period decrease.  In summary for all three known period change mechanisms for V1017 Sgr, the sum of all three known effects will result in a small {\it increase} of the period across the eruption.

This is a substantial problem for nova theorists.  But we have found the exact same problem for DQ Her, RR Pic, HR Del, and QZ Aur, and these cases are all with very high confidence levels that the periods {\it decreased} across their CN eruptions.  So V1017 Sgr is simply the fifth case of a confidently measured period decrease.  We have still not even heard of any vague idea expressed to explain the large period decreases.  This is now a strong challenge to the completeness of nova theories.

The large period decrease provides a larger challenge for the theory of CVs in that this unknown mechanism is shortening the orbital period at a faster rate than the two known mechanisms (gravitational radiation and magnetic breaking), so this unknown mechanism is dominating the evolution of four-out-of-six CN systems.  That is, for these four systems, the period changes by $\sim$$\frac{1}{3000}$ for each eruption, so it will only take $\sim$3000 eruptions before a simple extrapolation has the period going to zero.  While such an extrapolation will not be exact, it does set the time scale for the lifetime of the CV.  For V1017 Sgr, given the crudely known $M_{WD}$ and eruption amplitude, we can estimate that the inter-eruption time interval is of order $10^4$ years (Yaron et al. 2005).  So V1017 Sgr apparently has a lifetime of order $10^{7.5}$ years as driven by this unknown mechanism.  This time scale is greatly faster than the evolutionary time scale for the red giant companion star.  This fast lifetime has nothing to do with the evolved nature of the companion star, because RR Pic, HR Del, and QZ Aur are ordinary CNe with low mass companions near the main sequence.  For now, the unknown mechanism is dominating the evolution of at least the majority of CN systems, its applicability to other systems is unknown, and the implications for evolution have not been thought out.

The period decrease of V1017 Sgr provides a counter-example to the basic mechanism of the Hibernation model (Shara et al. 1986).  That is, the Hibernation model requires that the companion star separates slightly from the white dwarf, with this increased orbital period driving the entire Hibernation phenomenon.  Furthermore, Hibernation has the post-eruption nova slowly fading over the decades, as the irradiation effects diminish, with the system dimming into a hibernating state.  But V1017 Sgr fails both required predictions of Hibernation, as its period decreases and its brightness is not fading.  So V1017 Sgr provides a proof that Hibernation does not work, at least for this one nova system.

With a relatively large period decrease across the 1919 CN eruption, the size of the Roche lobe will decrease.  Suddenly the point for spilling matter from the companion star is pushed deeper into the stellar atmosphere.  With this, the accretion rate should increase suddenly across the eruption.  We can calculate the decrease in the Roche lobe associated with the period change.  This will be a weak function of the changing masses of the white dwarf and the companion star, as well as of the mass of the ejected shell.  For $10^{-6}$ M$_{\odot}$ ejected, $M_{WD}$=1.1 M$_{\odot}$, and $M_{comp}$=0.60 M$_{\odot}$, the size of the companion star's Roche lobe shrinks by 610 km.  This should be compared against the atmospheric scale height of the giant star's atmosphere.  For normal composition, 5.3 R$_{\odot}$ radius, and a surface temperature of 5200 K, the atmospheric scale height is 5200 km.  By moving the edge of the Roche lobe 610 km deeper in such an atmosphere, the density of the atmosphere at the Roche surface will increase by 17\%.  So the strong prediction is that for the observed period change, there should be an increase in the accretion rate, and hence in the accretion disk brightness by 13\% from before 1919 to after the end of the eruption.  This corresponds to close to 0.13 mag change.  This is within the variations displayed in Figure 4.  With the companion star dominating the optical light, even in the B band, the expected systematic change in magnitude due to the period change will be greatly smaller than 0.13 mag.  Thus, we conclude that the expected change in brightness across the 1919 CN event from this mechanism should be too small to measure.

\section{Conclusions}

	V1017 Sgr is a nearly unique CV that displays both CN and DN eruptions.  The only previous work has been scattered photometric measures, some spectra showing absorption lines for a G5 III star, and nine points in a radial velocity curve.  Sekiguchi's reported period of 5.714 days was acknowledged to be dubious and ambiguous.  And no one had ever followed up to test this period either photometrically or spectroscopically.  In this paper we report large numbers of new photometric measures aimed at testing Sekiguchi's claimed period.
	
	For this, we have collected 2896 magnitudes (plus 53 from the literature) in the UBVRIJHKL bands stretching from 1897 to 2016.  This now represents a nearly complete set of all photometry of V1017 Sgr.  
	
	We find that the SED is well represented by a 5200 K blackbody, which is the giant companion star dominating over the accretion disk.  The 1919 event was certainly a S-class CN, with the peak some weeks before the discovery on 11 March 1919 (as it was near solar conjunction at the time) with a likely peak magnitude of 4.5$\pm$1.0 (certainly between 6.4 and 2.0), for a total amplitude of 10.5 mag or so. The nova system showed a pre-eruption rise of 0.6 mag from 1907.5 to 1916.5, but it is problematic as to whether this is causally connected to the CN event.   Three DN events were seen in 1901, 1973, and 1991, all with effectively identical light curves.  Their total durations were 0.5 years each,  the amplitude was close to 3 mags, while the rise and fall was symmetric and roughly triangular in shape.  V1017 Sgr is the {\it only} known CN that displays DN events {\it before} the CN event.  The quiescent level after the CN event (from 1923 to 2016) shows no systematic decline, but rather displays 0.3-0.6 mag variations on the decadal time scale.  We do not understand how the system can vary by so much in quiescence (with equal amplitudes in BVRI colors) when the light is dominated by a presumably constant star of fixed radius.
	
	For the post-eruption magnitudes, we find a  highly significant photometric periodicity 2.893145$\pm$0.000016 days, with a small $\dot{P}$.  Given the dominance of the companion light so that the photometric modulation must be dominated by the ellipsoidal effect, and the Sekiguchi radial velocities, the true orbital period must be twice the photometric period.  When folded on the orbital period, the light curve alternates between deep and shallow minima, with the unusually-large accretion disk eclipsing the companion star during the deeper minima.  With the orbital period being twice the photometric period, we see that Sekiguchi got the correct period, even if with poor accuracy.
	
	This orbital period is one of the longest amongst all known CVs.  This requires that the companion star have a size of close to 5.3 R$_{\odot}$, and certainly having evolved off the main sequence.  This implies that the current accretion is being driven by the evolutionary expansion of the companion star, and not by some angular momentum loss mechanism.  This also requires that the accretion disk around the white dwarf be exceptionally large.  In this case, the outer regions of the disk can be cool enough for the DN instability to occur, despite the relatively high accretion rate that is needed for the CN event.  So the very long orbital period explains why V1017 Sgr is the only CV with a DN event before a CN eruption.

	With the orbital period and the SED, we derive a blackbody distance to the companion star to be 1240$\pm$200 parsecs.  This relatively nearby distance plus the relatively high luminosity giant companion star gives the reason why V1017 Sgr is one of the brightest ex-nova in the sky.
	
	We have 15 useable magnitudes from before the 1919 eruption, and these give a periodicity with a good looking folded light curve, and the best fit connects with the post-eruption ephemeris at the time of the CN event.  The pre-eruption period is different from the post-eruption period at the 5.7-sigma confidence level.  The period decreased by -273$\pm$61 parts-per-million.  This large period decrease is impossible to understand with any combination of the three known mechanisms for period change across a CN event.

	~
	
	~
	
	We thank the many observers who have contributed to the AAVSO International Database for their unique and valuable coverage of V1017 Sgr.  AUL acknowledges support from NSF grant AST 0803158.



\bsp	
\label{lastpage}

\begin{thebibliography}{99}
\bibitem[\protect\citeauthoryear{Cardelli et al.}{1989}]{Cardelli et al. 1989}
Cardelli, J. A., Clayton, G. C, \& Mathis, J. S., 1989, ApJ, 345, 245
\bibitem[\protect\citeauthoryear{Collazzi et al.}{2009}]{Collazzi et al. 2009}
Collazzi, A. C., Schaefer, B. E., Xiao, L., Pagnotta, A., Kroll, P., \& Lochel, K., 2009, AJ, 138, 1846
\bibitem[\protect\citeauthoryear{Downes \& Duerbeck}{2000}]{Downes and Durebeck 2000}
Downes, R. A. \& Durebeck, H. W., 2000, AJ, 120, 2007
\bibitem[\protect\citeauthoryear{Frank et al.}{2002}]{Frank et al. 2002}
Frank, J., King, A., \& Raine, D., 2002, Accretion Power in Astrophysics, 3rd ed. (Cambridge: Cambridge University Press
\bibitem[\protect\citeauthoryear{Harrison et al.}{1993}]{Harrison et al. 1993}
Harrison, T. E., Johnson, J. J., \& Spyromilio, J., 1993, AJ, 105, 320
\bibitem[\protect\citeauthoryear{Kasliwal et al.}{2011}]{Kasliwal 2011}
Kasliwal, M. M., Cenko, S. B., Kulkarni, S. R., Ofek, E. O., Quimby, R., \& Rau, A., 2011, ApJ, 735, 94
\bibitem[\protect\citeauthoryear{Kraft}{1964}]{Kraft 1964}
Kraft, R. P., 1964, ApJ, 139, 457
\bibitem[\protect\citeauthoryear{Landolt}{1975}]{Landolt 1975}
Landolt, A. U., 1975, PASP, 87, 265
\bibitem[\protect\citeauthoryear{Landolt}{2016}]{Landolt 2016}
Landolt, A. U., 2016, JAAVSO, 44, 45
\bibitem[\protect\citeauthoryear{Livio}{1991}]{Livio 1991}
Livio, M., 1991, ApJLett, 369, L5
\bibitem[\protect\citeauthoryear{Martin et al.}{2011}]{Martin et al. 2011}
Martin, R. G., Livio, M., \& Schaefer, B. E., 2011, MNRAS, 415, 1907
\bibitem[\protect\citeauthoryear{McLaughlin}{1946}]{McLaughlin 1946}
McLaughlin, D. B., 1946, PASP, 58, 46
\bibitem[\protect\citeauthoryear{Mroz et al.}{2016}]{Mroz et al. 2016}
Mroz, P., Udalski, A., Pietrukowics, P., Szymanski, M.K., et al., 2016, Nature, 537, 649
\bibitem[\protect\citeauthoryear{Mumford}{1971}]{Mumford 1971}
Mumford, G. S., 1971, ApJ, 165, 369
\bibitem[\protect\citeauthoryear{Patterson et al.}{2013}]{Patterson et al. 2013}
Patterson, J., Uthas, H., Kemp, J., de Miguel, E., et al., 2013, MNRAS, 434, 1902
\bibitem[\protect\citeauthoryear{Prialnik \& Shara}{1986}]{Prialnik and Shara 1986}
Prialnik, D. \& Shara, M. M., 1986, ApJ, 311, 172
\bibitem[\protect\citeauthoryear{Schaefer}{2010}]{Schaefer 2010}
Schaefer, B. E., 2010, ApJSuppl, 187, 275
\bibitem[\protect\citeauthoryear{Schaefer}{2011}]{Schaefer 2011}
Schaefer, B. E., 2011, ApJ, 742, 112
\bibitem[\protect\citeauthoryear{Schaefer \& Collazzi}{2010}]{Schaefer and Collazzi 2010}
Schaefer, B. E. \& Collazzi, A. C., 2010, AJ, 139, 1831
\bibitem[\protect\citeauthoryear{Schaefer \& Patterson}{1983}]{Schaefer and Patterson 1983}
Schaefer, B. E. \& Patterson, J., 1983, ApJ, 268, 710
\bibitem[\protect\citeauthoryear{Schaefer et al.}{2013}]{Schaefer et al. 2013}
Schaefer, B. E., Landolt, A. U., Linnolt, M., Stubbings, R., et al., 2013, ApJ, 773, 55
\bibitem[\protect\citeauthoryear{Schlafly \& Finkbeiner}{2011}]{Schlafly and Finkbeiner 2011}
Schlafly, E. F. \& Finkbeiner, D. P., 2011, ApJ, 737, 103
\bibitem[\protect\citeauthoryear{Shara et al.}{1986}]{Shara et al. 1986}
Shara, M. M., Livio, M., Moffat, A. F. J., \& Orio, M., 1986, ApJ, 311, 163
\bibitem[\protect\citeauthoryear{Sekiguchi}{1992}]{Sekiguchi 1992}
Sekiguchi, K., 1992, Nature, 358, 563
\bibitem[\protect\citeauthoryear{Strope et al.}{2010}]{Strope et al. 2010}
Strope, R. J., Schaefer, B. E., \& Henden, A. A., 2010, AJ, 140, 34
\bibitem[\protect\citeauthoryear{Vidal and Rodgers}{1974}]{Vidal and Rodgers 1974}
Vidal, N. V. \& Rodgers, A. W., 1974, PASP, 86, 26
\bibitem[\protect\citeauthoryear{Walker}{1977}]{Walker 1977}
Walker, A. R., 1977, MNRAS, 179, 587
\bibitem[\protect\citeauthoryear{Webbink et al.}{1987}]{Webbink et al. 1987}
Webbink, R. F., Livio, M, Truran, J. W. \& Orio, M., 1987, ApJ, 314, 653
\bibitem[\protect\citeauthoryear{Woods}{1919}]{Woods 1919}
Woods, I., 1919, Harvard College Observatory Bulletin, 693, 1
\bibitem[\protect\citeauthoryear{Yaron et al.}{2005}]{Yaron et al. 2005}
Yaron, O., Prialnik, D., Shara, M. M. \& Kovetz, A., 2005, ApJ, 623, 398

\end{thebibliography}
\end{document}